\documentclass[a4paper]{jpconf}
\usepackage{graphicx}

\usepackage{bbm}
\usepackage[utf8]{inputenc}
\usepackage{placeins}
\usepackage{amssymb,amsmath,amsfonts}
\usepackage[colorlinks,linkcolor=purple,citecolor=teal]{hyperref}
\usepackage{dsfont}
\usepackage{color}
\usepackage{graphicx}
\usepackage{hyphenat}
\usepackage{wrapfig}
\usepackage{empheq}
\usepackage{textcomp}
\usepackage[caption=false]{subfig}
\usepackage{rotating}
\usepackage{wrapfig}
\usepackage{pdfpages}
\usepackage{verbatim}
\usepackage{setspace}
\usepackage{array}
\usepackage{caption}
\usepackage[pdf]{pstricks}
\usepackage{upgreek}
\usepackage{cmll}
\usepackage{latexsym}
\usepackage{braket}
\usepackage{cite}
\usepackage{epsfig}
\usepackage[left=2.4cm,top=3.3cm,right=2.4cm,bottom=3.3cm,bindingoffset=0cm]{geometry}
\usepackage[titletoc,page]{appendix}
\usepackage{multicol}
\usepackage{youngtab}
\usepackage[normalem]{ulem}     

\newcommand{\beq}{\begin{eqnarray}}
\newcommand{\eeq}{\end{eqnarray}}
\newcommand{\bea}{\begin{eqnarray}}
\newcommand{\bqa}{\begin{eqnarray}}
\newcommand{\eea}{\end{eqnarray}}
\newcommand{\be}{\begin{equation}}
\newcommand{\ee}{\end{equation}}
 
\newcommand{\im}{\mathrm{i}}
\newcommand{\calR}{\mathcal{R}}
\newcommand{\rmc}{\mathrm{c}}

\newcommand{\rmF}{\mathrm{F}}
\newcommand{\rmL}{\mathrm{L}}

\newcommand{\rmS}{\mathrm{S}}
\def\brc{\langle}
\def\ckt{\rangle}
\def\const{{\rm const}}
\def\de{\partial}

\def\nn{\nonumber}

\def\Tr{\qopname\relax o{Tr}}

\numberwithin{equation}{section}

\def\bxi{{\bar{\xi}} }
\def\su{$ \phantom{{{{{\yng(1)}}}}}\!\!\!\!\!\!\!\!$}
\def\sbuu{$\phantom{{{{\bar{\yng(1,1)}}}}}\!\!\!\!\!\!$}

\def\sbbu{ $\phantom{{{\bar{\bar{\yng(1)}}}}}\!\!\!\!\!\!\!\!$ }

\numberwithin{equation}{section}

\begin{document}
\title{Dynamics of strongly-coupled chiral gauge theories}

\author{Stefano Bolognesi$^{1,2}$, Kenichi Konishi$^{1,2}$ and Andrea Luzio$^{2,3}$}

\address{$^1$ Department of Physics E. Fermi, University of Pisa,
Largo Pontecorvo, 3, Ed. C, 56127 Pisa, Italy}

\address{$^2$ INFN, Sezione di Pisa, Largo Pontecorvo,
3, Ed. C, 56127 Pisa, Italy}

\address{$^3$ Scuola Normale Superiore, Piazza dei Cavalieri,
7, 56127 Pisa, Italy}

\ead{stefano.bolognesi@unipi.it, kenichi.konishi@unipi.it, andrea.luzio@sns.it}

\begin{abstract}

We study the dynamics of  $SU(N)$ chiral gauge theories with massless fermions belonging to various combinations of the symmetric, antisymmetric or fundamental representations. 
We limit ourselves to the gauge-anomaly-free and asymptotically  free systems. These theories have a global symmetry group with the associated   't Hooft anomaly-matching conditions severely limiting  the  possible RG flows.  Recent developments on the applications of the generalized symmetries and the stronger requirement of the matching of the mixed anomalies also give further  indications on the possible  IR dynamics. 
In vectorlike theories such as the quantum chromodynamics  (QCD),  gauge-invariant ``quark-antiquark" condensates form and characterize the IR dynamics, and the anomaly matching involves the Nambu-Goldstone (NG)  bosons.  In some other special cases, such as the Bars-Yankielowicz (BY)  or Georgi-Glashow (GG) models, a hypothetical solution was proposed in the literature, with no global symmetry breaking and with some simple set of composite massless fermions  saturating all  the anomalies.  For the BY and GG systems, actually,  a more plausible candidate for their IR physics is  the dynamical Higgs phase, with a few simple bi-fermion color-flavor locked condensates, breaking the color and flavor symmetries,  partially or totally. Remarkably, the  't Hooft anomaly-matching (and generalized anomaly-matching) conditions are automatically  satisfied in this phase.
Another interesting possibility, occurring in some chiral gauge theories, is dynamical Abelianization,  familiar from ${\cal N}=2$ supersymmetric gauge theories. 
We explore here even more general types of possible IR phases than the ones mentioned above,  for wider classes of models.  With the help of  large-N  arguments we look for IR free theories, whereas  the  MAC (maximal attractive channel) criterion might suggest some simple bi-fermion condensates characterizing the IR dynamics of the systems. 
In many cases the low-energy effective theories are found to be described by quiver-like gauge theories, some of the (nonAbelian)  gauge groups are infrared-free while some others might be asymptotically free.

\end{abstract}

\newpage

\section{Introduction}

It is the purpose of this review work to discuss a wide class of strongly-coupled $SU(N)$ chiral gauge theories and their possible IR physics,  based on the work  \cite{BKS,BK,BKL1,BKL2,BKL4,BKL5,BKLReview,BKLDA} done in the last several years.  
In spite of many years of efforts \cite{Raby:1979my,Appelquist:1999vs,Appelquist:2000qg,Bars:1981se,Eichten:1985fs,Geng:1986xh,Goity:1985tf,Ibanez:1991hv,Shi:2015fna,Shi:2015baa,Shifman:2008cx,Dimopoulos:1980hn}, and in spite of important possible applications in the context of realistic model building beyond the standard model of the fundamental interactions,  
 our understanding of the dynamics of {\it  chiral} gauge theories is still quite unsatisfactory today.  Such a consideration motivated us to study  wide classes of 
$SU(N)$ gauge theories with massless matter fermions in various possible representations, systematically.  The main tool  of the analysis is the constraints following from the requirement that the global symmetries of the systems be correctly reflected as the system RG-flows towards IR, and as the interactions become strong.  Of particular importance in this context is the consideration of the 't Hooft anomalies \cite{THooft}
(the obstructions in gauging the global symmetries by introducing external, arbitrarily weakly coupled gauge fields), and  their consequences in the RG flow.  These anomalies must be faithfully reproduced by the  degrees of  freedom in the low-energy effective theory,    either by massless composite fermions for unbroken symmetries (the 't Hooft anomaly matching constraints) or by interactions involving the Nambu-Goldstone (NG) bosons if some of the global symmetry is spontaneously broken,  or in some cases  by both.

As is well known, the choice of the fermion representations is restricted by the gauge anomaly-free  conditions.  Another condition we impose  is that  the system is asymptotically free.  In our studies below, we further restrict possible reducible representations for the fermions to those involving  the fundamental representation, the rank-two symmetric or antisymmetric representations,  and their conjugates, only.  As we will see they cover already  very rich variety of systems of physical interest. 

An aspect of these constraints that various symmetries impose on the RG flow (physics in the IR),  which is perhaps less familiar,  is  the low-energy manifestations of the classical $U(1)$ symmetry, made anomalous by the strong gauge interactions (``the strong anomaly").   The so-called $U(1)_A$ problem and its solution in quantum chromodynamics
(QCD) - which is the consequence of such a strong anomaly  -  is well-known   \cite{Witten:1980sp,Witten:1979vv,DiVecchia:1980yfw,Rosenzweig:1979ay,Kawarabayashi:1980dp,Nath:1979ik},   
but for some reason this particular aspect of the symmetry consideration has been applied  in the context of chiral gauge theories only recently \cite{BKL5}  (see however \cite{Veneziano:1981yz}).

What makes these problems highly nontrivial and interesting is the fact 
 that the strong $SU(N)$ gauge interactions themselves can manifest in different phases  at low energies, depending on the matter fermions present,  such as  confinement, Higgs, Coulomb, or still other phases. To understand the interplay between the types of massless matter fermions and these phases of the system under study, is one of the central themes of this work.

Recently the ideas of generalized, higher-form symmetries and their gauging have been applied 
to gain deeper insights into these problems  \cite{Shimizu:2017asf,AhaSeiTac,ShiYon,Tanizaki:2018wtg,GKSW,GKKS,TanKikMisSak,Anber:2018iof}.   One of the key tools, which leads to many interesting consequences,  
 is the  ${\mathbbm Z}_N $ center symmetry
 of $SU(N)$ theories. Although the idea of the center symmetry itself is a familiar one,\footnote{A precursor of the ideas is indeed  the center symmetry ${\mathbbm Z}_N$ in Euclidean  $SU(N)$ Yang-Mills theory at finite temperature, which acts on the Polyakov loop. The unbroken (or broken) center symmetry  by the VEV of the Polyakov loop, is a criterion of confinement (or de-confinement) phase \cite{Polyakov}.}  it becomes more powerful when combined with the idea of ``gauging" 
such a discrete center symmetry \cite{Shimizu:2017asf,AhaSeiTac,ShiYon,Tanizaki:2018wtg,GKSW,GKKS,TanKikMisSak,Anber:2018iof}.  In some cases,  this leads to mixed  ([0-form]-[1-form]) 't Hooft anomalies;  the matching of these new types of anomalies imposes stronger constraints  
 on the possible infrared dynamics of the system, as compared with the conventional 't Hooft anomaly matching conditions.

 This work is basically a review of our previous results \cite{BKS,BK,BKL1,BKL2,BKL4,BKL5,BKLReview,BKLDA},     but  a special emphasis will be  put on an exploration of more general, new types of phases and low-energy effective actions,  than those already discussed in literature or  by ourselves. Keeping such an aim in mind,
 the discussions on the generalized symmetries and applications of the mixed anomalies will be left to another review work  \cite{BKLReview}.
 The present work is organized as follows. 
 In Sec.~\ref{sec:class}  we discuss the classes of models to be analyzed below.   In the following sections, Sec.~\ref{Hypo}, Sec.~\ref{DHiggs}, and in Sec.~\ref{DA},  we review the three possible different types of phases which may be realized, depending on the matter content of our theories,   i.e., a hypothetical confining flavor-symmetric vacuum,  dynamical Higgs phase, and dynamical Abelianization, respectively.  In Sec.~\ref{new}   we explore  other possible phases, which are generalizations of the dynamical Higgs/Abelianization  cases discussed in     Sec.~\ref{DHiggs}, and in Sec.~\ref{DA}.   
 The content of Sec.~\ref{new}, which is the tentative study of more general IR dynamics,   is mostly new.     
 Sec.~\ref{StrAn} reviews  the implications of the strong anomaly  to possible dynamical scenarios in the IR in some simple chiral gauge theories.    Conclusive remarks are in Sec.~\ref{Concl}.

\section{Models} 
\label{sec:class}

In this paper we  shall focus on asymptotically-free $SU(N)$ gauge theories with chiral fermions. We restrict further our playground to the class of  theories which possess a large N limit.  
The last requirement  imposes that the matter content is restricted to  a combination of a few  irreducible representations (irreps) of $SU(N)$,\footnote{All  fermions are taken to be left-handed.  }
\beq 
&& N_{\psi}\, \yng(2) \oplus   N_{\tilde{\psi}} \, \bar{ \yng(2)} \oplus N_{\tilde{\chi}} \,\yng(1,1) \oplus   N_{\chi} \,\bar{ \yng(1,1)} \oplus  N_{\tilde{\eta}}\,{   {{\yng(1)}}}\; \oplus   N_{{\eta}}   \,{\bar   {{\yng(1)}}}\; \oplus N_{\lambda} \!\!\!\!\! \!\!\begin{array}{c}\;\;\;\;\yng(2,1) \\ \vdots \\   \yng(1,1)\end{array} \nn \\
&& \qquad   \  \psi \qquad \quad \ \  \,  \tilde\psi  \qquad \quad \ \  \tilde\chi \qquad \ \  \chi \qquad \quad   \tilde\eta \qquad \quad \ \, \eta \qquad  \quad   \  \lambda \ ,
\label{general1}
\eeq
inclusion of larger irreps will render the theory IR-free at large N.

Not all the choices of $N_{\psi},\; N_{\tilde{\psi}},\; N_{\tilde{\chi}},\; N_{\chi}, \;N_{\tilde{\eta}},\; N_{{\eta}},\; N_{{\lambda}}$ are permitted: gauge anomaly cancellation imposes that
\beq
(N_{\psi} -   N_{\tilde{\psi}}) (N+4) + (N_{\tilde{\chi}} -  N_{\chi}) (N-4) +  (N_{\tilde{\eta}} -  N_{{\eta}} ) N = 0\;,
\label{general2}
\eeq
which reduces our parameter spaces from seven  to six integers. Asymptotic freedom\footnote{In general these theories become strongly coupled in IR, and develop
a dynamically generated energy scale, $\Lambda$. Some of them might however flow to an interacting CFT, as can be suggested by the analysis which takes into account the higher coefficients of the beta function. }
\be   
b_0= 11N -  (N_{\psi} + N_{\tilde{\psi}}) (N+2) - (N_{\tilde{\chi}} +  N_{\chi}) (N-2) -   (N_{\tilde{\eta}} + N_{{\eta}} )  - 2  N   N_{\lambda}  > 0  \;.  \label{betafn}
\ee
This gives an inequality on the parameter space, thus it doesn't reduce the "dimensionality" of the parameter spaces, but renders  the number of theories we can consider, for fixed $N$, finite.

Before giving a convenient parametrization of this six-dimensional set of theories, let us see that it contains many interesting theories. 

Clearly, it contains all the vector-like theories that have a ('t Hooft) large-N limit. In particular, for $N_{\tilde{\eta}} = N_{{\eta}}$ and setting all rest $N$s to zero one obtains ordinary QCD. Similarly, by allowing only $N_{\psi}=N_{\tilde{\psi}}$ ($N_{\tilde{\chi}}=N_{\chi}$) to be non-zero one obtains (S)QCD ((A)QCD), in which the ``quarks" are in the second-rank tensor representations of $SU(N)$.  Lastly, by allowing only $N_{adj.}\neq 0$, one obtains adjoint-QCD.

This family contains also many examples of chiral gauge theories.
For example, taking $N_{S}=1$, $N_{{\eta}}=(N+4)$ and setting all the others $N$s to zero one obtains an $SU(N)$ gauge theory  with left-handed fermions in the reducible, complex representation, 
\beq       
\yng(2) \oplus   (N+4) \, {\bar{{\yng(1)}}} \ .   \label{psieta} 
\eeq
This is the simplest of the so-called Bars-Yankielowicz (BY)  models \cite{Bars:1981se}.  We call it ``$\psi\eta$"  model,  for short:  its IR phase has been studied extensively, see \cite{BKL1,BKL2,BKL4, BKL5,BKLReview}. 

Another simple  example is  $N_{\tilde{\chi}}=1$, $N_{{\eta}}=(N-4)$, $N\ge 5$, i.e.
\be       \yng(1,1) \oplus   (N-4) \,{\bar   {{\yng(1)}}}\, .\ee
This is the simplest  Georgi-Glashow  (GG)  model. We will refer to it as   ``$\tilde\chi\eta$"  model.

A still another simple model can be constructed by taking $N_{\psi}=1,\;N_{\chi}=1,\;N_{{\eta}}=8$,   i.e.,  
\be     \yng(2) \oplus  {\bar  {\yng(1,1)}}  \oplus   8  \   {\bar  {\yng(1)}}\;.  \label{psichieta}
\ee
This theory that we called $\psi\chi\eta$ model, has been studied earlier in \cite{Appelquist:2000qg,Goity:1985tf,Eichten:1985fs}   and more recently, in   \cite{BKS,BK,BKLDA,Sheu:2022odl}.

As introduced in 
\cite{BK} we consider the 
$(N_{\psi},N_{\chi})$ models. They are minimal models of the type (\ref{general1}) that do not have any sub-vectorial sector \footnote{
$N_{\psi}$, when positive, stands for the number of $\tiny{\yng(2)}$ and when negative stands for the number of $\tiny{\bar{\yng(2)}}$.  $N_{\chi}$, when positive, stands for the number of $\tiny{\bar{\yng(1,1)}}$ and when negative stands for the number of $\tiny{ \yng(1,1)}$. We choose the values  $(N_{\psi},N_{\chi})$ and then we add the minimal number of fundamental, or anti-fundamental to 
  cancel the gauge anomaly. }.
  There are three categories of models.
  We denote them as type I, II  and III. For the first one:
\bea
&& \  {\rm type   \ (I) }\,, \ \  N_{\psi} \geq N_{\chi}\geq 0 \ , \nn  \\ 
&& \    N_{\psi}  \, \yng(2)   \oplus  N_{\chi} \, {\bar  {\yng(1,1)} }   \oplus \big(N_{\psi}(N+4) -N_{\chi}(N-4)  \big)\, \,  {\bar  {\yng(1)} }  \ ,   \nn  \\ 
&& \  
\ b_0 = (11 - 2N_{\psi}) N -6 N_{\psi} -2 N_{\chi}\ .  \phantom{ {{\yng(1,1)} } }     \label{typeI}
\eea 
For asymptotic large $N$, $N_{\psi}$ can go up to $5$ while retaining asymptotic freedom (AF).
For the second class:
\bea 
&&  \ {\rm type   \ (II) } \,, \ \  N_{\chi}  > N_{\psi} \geq 0   \ , \nn  \\ 
&& \    N_{\psi}\,    \yng(2)   \oplus   N_{\chi} \,  {\bar  {\yng(1,1)} }   \oplus   \big(N_{\chi}(N-4) -N_{\psi}(N+4)   \big)  \, {  {\yng(1)} }  \ ,      \nn  \\ 
&& \  
b_0 = (11 - 2N_{\chi}) N +2 N_{\psi} +6 N_{\chi} \ .  \phantom{ {{\yng(1,1)} } }  
\label{typeII}
\eea 
For   large $N$, $N_{\chi} \leq 5$ to have AF.
For the third class:
\bea 
&&  \  {\rm type   \ (III) }\,,   \ \   N_{\psi} \geq 0 \geq N_{\chi} \ , \nn \\ 
&& \     N_{\psi}  \,  \yng(2)   \oplus   (-N_{\chi}) \,  {  {\yng(1,1)} }   \oplus   \big(N_{\psi}(N+4) -N_{\chi}(N-4)   \big)  \,  {\bar  {\yng(1)} }  \ ,     \nn  \\ 
&& \  
\ b_0 = (11 - 2(N_{\psi}-N_{\chi})) N - 6  N_{\psi} - 6 N_{\chi} \ . \phantom{ {{\yng(1,1)} } }    \label{typeIII}
\eea  
For   large $N$, $N_{\psi} - N_{\chi} \leq 5$ to have AF.

\begin{figure}[h!]
\begin{center}
\includegraphics[width=4in]{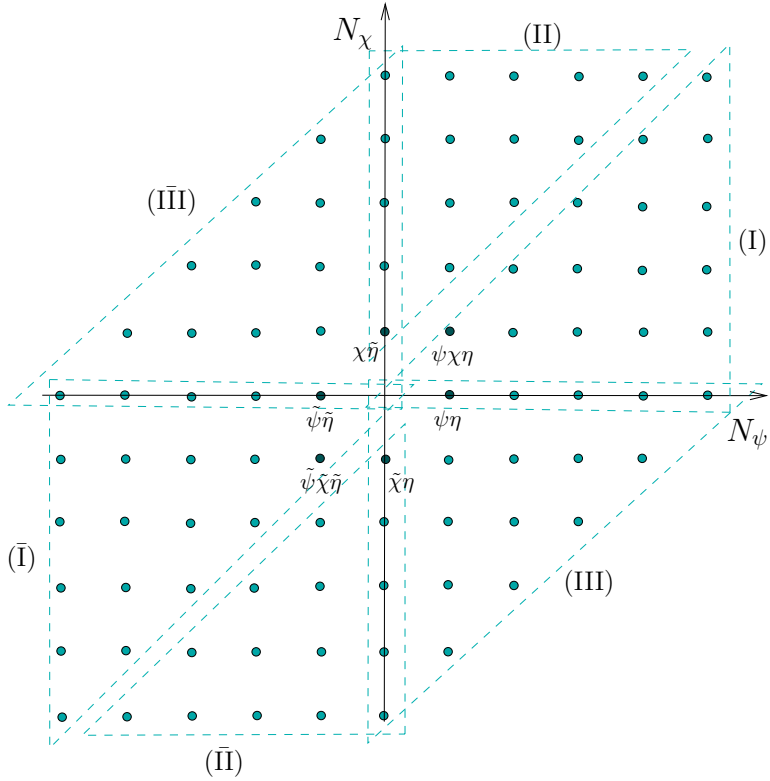}
\caption{\small All the possible  $(N_{\psi},N_{\chi})$ models that are AF at large $N$. }
\label{theories}
\end{center}
\end{figure}

\begin{figure}[h!]
\begin{center}
\includegraphics[width=6in]{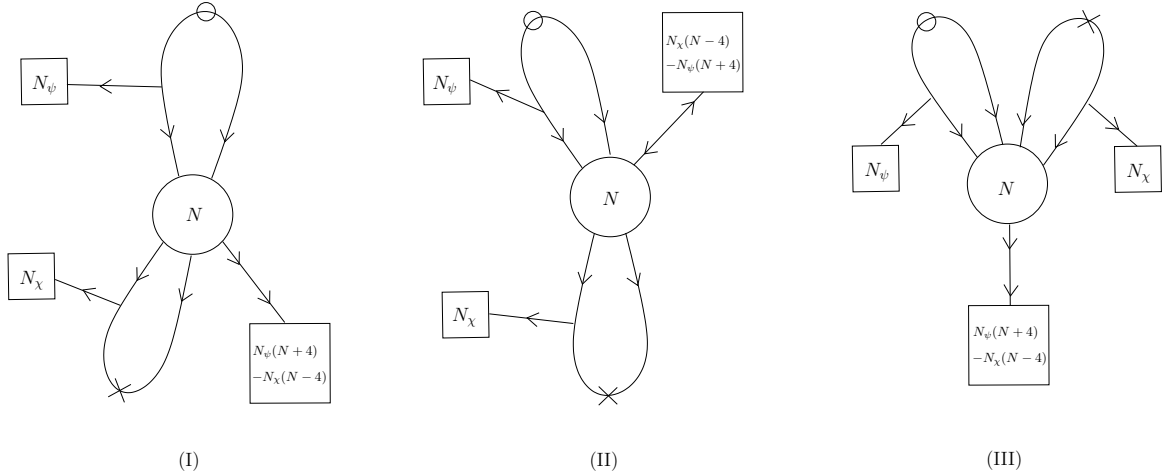}
\caption{\small Quiver diagram   to $(N_{\psi},N_{\chi})$ models models of type (I), (II) and (III). There are other $U(1)$   global charges which are not shown in the diagram. }
\label{quiv}
\end{center}
\end{figure}

Among the $(N_{\psi},N_{\chi})$ models $(1,0)$ correspond to the $\psi \eta$ model, $(1,1)$ corresponds to the $\psi\chi \eta$ model and
$(0,1)$ corresponds to the $ \chi \tilde \eta$  model. These three types, plus their complex conjugate $(\bar{\rm{I}})$, $(\bar{\rm{II}})$ $(\bar{\rm{III}})$ with  $N_{\psi} \leq 0$ are shown  in Figure \ref{theories} in the $(N_{\psi},N_{\chi})$ plane . The boundaries are set by the asymptotic-freedom  requirement   $
b_0 >0$  evaluated at large $N$.  
We introduce a quiver notation which may be useful to visualize the theories and their differences.
We use the following notation for the quiver diagram: circles with a number inside $n$ represent a gauge group $SU(n)$, squares with a number $m$ represent a global symmetry $SU(m)$, fermions are lines connecting the groups, arrows on the line indicate if is fundamental (ingoing) or antifundamental (outgoing), 
little ``o'' or ``x'' within a line indicates if the ends are symmetric or anti-symmetric.
See figure \ref{quiv} for the diagrams of the $(N_{\psi},N_{\chi})$ models of types (I), (II) and (III).

 The complete six-parameter set of theories of  (\ref{general1}), (\ref{general2}) can be then parametrize by a $(N_{\psi},N_{\chi})$ model plus various vectorial copies of   QCD,  (S)QCD,  (A)QCD or adjoint QCD.

As an example, starting from the $\psi\eta$ model one can obtain  all the Bars-Yankielowicz  (BY)   models  
\be       \yng(2) \oplus   (N+4+p) \,{\bar   {{\yng(1)}}}\;\oplus   p \,{   {{\yng(1)}}}\; , \label{psietabaretadef}
\ee
where the number of the extra fundamental pairs  $p$ is limited by $\frac{9}{2}N-3$ before asymptotic freedom is lost.  

Also the Georgi-Glashow models 
\be       \yng(1,1) \oplus   (N-4+p) \,{\bar   {{\yng(1)}}}\;\oplus   p \,{   {{\yng(1)}}}\;. \label{def:GG}
\ee 
can be constructed starting from the $(N_\psi=0, N_\chi=-1)$ model (the $\tilde{ \chi} \eta$ model), by adding $p$ Dirac fundamental fermions.
Here  $p$ will be assumed to be less than $\frac{9}{2}N+3$ so as to  maintain AF.

These theories can have various types of gauge-invariant operators. Whenever there is a vectorial part like 
QCD,  (S)QCD or  (A)QCD, there are gauge invariant bi-fermions $\eta\tilde{\eta} $, $\tilde{\psi}\psi$ or $\chi \tilde{\chi}$.  These operators can be scalars, and we expect them to condense as they are usually in the strongest possible attractive channel, and also because they do in normal QCD. In the $(N_{\psi},N_{\chi})$ models it is not possible to have bi-fermion  gauge-invariant operators.   For type (I) and (II) theories 
we have an adjoint bi-fermion operator $\psi\chi$ which can be made gauge invariant with the addition of some gluon operator and then traced appropriately. These operators can also be scalars and in principle they could condense, although we do not consider this possibility. There are 
in general three-fermion gauge invariant objects, for example $\psi\eta\eta$, $\bar{\chi}\eta\eta$ in type (I), $\chi\tilde{\eta}\tilde{\eta}$, $\bar{\psi}\tilde{\eta}\tilde{\eta}$ in type (II) and $\psi\eta\eta$, $\tilde{\chi}\eta\eta$ in type (II). These can have a role in the IR as massless composite fermions.  There are four-fermion gauge-invariant operators $\bar{\eta}\bar{\psi}\psi\eta$ $\bar{\eta}\chi\bar{\chi}\eta$ or $  \psi \chi \psi \chi  $ in type (I) and similar for the others. These states might play a role   as mesonic states. These operators can also be scalar and, in principle,  they could condense, although we do not explore this possibility here.

The Large N limit of these theories contains a mixture of features some are typical of 't Hooft, Veneziano or (S,A)QCD-type,  large N limits.  ''Open-strings'' type states, like mesons $\eta\tilde{\eta}$ in QCD have a three-vertex interaction that scales like $g_s \sim \frac{1}{\sqrt{N}}$ at large-$N$.  The same happens for all open-strings type states like $\eta\psi\eta$, $\eta\tilde{\chi}\eta $, $\bar{\eta}\bar{\psi}\psi\eta$, ... .  ''Closed-strings'' type states, like mesons  $\tilde{\psi}\psi$ or $\chi \tilde{\chi}$ in (S,A)QCD have a three-vertex interaction that scales like $g_s \sim \frac{1}{N}$ at large-$N$. The same happens for all   states made of multiple $\psi$'s and $\chi$'s traced.
Large N argument suggests that the IR theory is weakly coupled for $N \to \infty$. We will look for IR free theories, but we should keep in mind that there is always the possibility of IR interacting fixed points, especially near the border of the diagram of Fig.\ref{theories}.

\begin{figure}[h!]
\begin{center}
\includegraphics[width=4in]{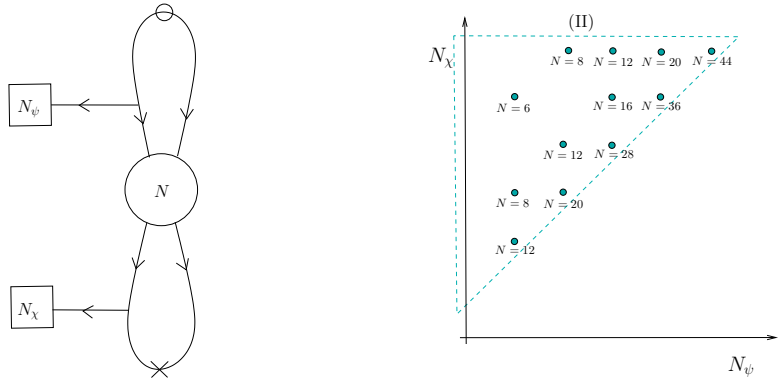}
\caption{\small    $(N_{\psi},N_{\chi})$   models without fundamentals. }
\label{nk}
\end{center}
\end{figure}

Inside this class there are models without fundamentals \cite{BKL1,Anber:2021iip}. (S)QCD, (A)QCD and adjoint QCD are  however vectorial. Chiral theories can be built with  fermions
   \be   \left( N_{\psi}=\frac{N-4}{k} \right)  \,    \yng(2)   \oplus \left( N_{\chi} =  \frac{N+4}{k}    \right)  \,    {\bar  {\yng(1,1)} }  \ ,   \label{basicnofund}
\ee
where $k$ is a common divisor of $N-4$ and $N+4$.  These examples are the first theories of type (II) (\ref{typeII}) where the number of fundamental is zero, $N_{\psi}(N+4) -N_{\chi}(N-4)=0$, they are represented in Figure \ref{nk}. 
 This type of theories cannot be scaled to large $N$ without fundamentals. On top of (\ref{basicnofund}) we can add multiple copies of $N_{\psi} $ (S)QCD,  $N_{\tilde{\chi}} $  (A)QCD and $N_{\lambda}$ adjoint QCD. An interesting feature of these models is that they have exact 
$\mathbb{Z}_2$ center symmetry for $N$ even. Another feature is that they do not have fermionic gauge-invariant operator when $N=4k$, in a sense they are 'purely bosonic' theories. 

 \subsection{Constraints for AF and infrared fixed-point conformal theory (CFT)}  

The requirement of AF  used in Table~\ref{theories}  to restrict the classes of theories, is actually subtler than that implied by the first coefficient of the beta function, $b_0$,  in (\ref{typeI})-(\ref{typeIII}).   A theory with  $b_0>0$ may actually flow into  a CFT,   as shown by Banks and Zaks \cite{BZ}  for  QCD  at large $N$ and $N_F$ near  
$N_F=   11 N / 2$,   and  by Seiberg \cite{SeibergDual}   for supersymmetric QCD  in the range  $  3 N /2 <    N_F  < 3 N $.

 We have examined whether or not some of the theories listed in   Table~\ref{theories}  (all with  $b_0>0$)  can flow into a CFT,  by taking into account the second coefficient
 of the beta function, at large $N$  \cite{Anber:2021iip,Zoller}.     The result is shown in the  Table below for type (I) and (II)
 \be
\alpha_{N_\psi, N_\chi}=\left(
\begin{array}{cccccc}
 -\tfrac{22 \pi }{17 N} & -\frac{24 \pi }{13 N} & -\frac{28 \pi }{5 N} & \frac{40 \pi }{19 N} & \frac{\pi }{2 N} & \frac{8 \pi }{77 N} \\
 -\frac{24 \pi }{13 N} & -\frac{2 \pi }{N} & -\frac{8 \pi }{N} & \frac{20 \pi }{11 N} & \frac{8 \pi }{17 N} & \frac{\pi }{10 N} \\
 -\frac{28 \pi }{5 N} & -\frac{8 \pi }{N} & -\frac{14 \pi }{N} & \frac{8 \pi }{5 N} & \frac{4 \pi }{9 N} & \frac{8 \pi }{83 N} \\
 \frac{40 \pi }{19 N} & \frac{20 \pi }{11 N} & \frac{8 \pi }{5 N} & \frac{10 \pi }{7 N} & \frac{8 \pi }{19 N} & \frac{4 \pi }{43 N} \\
 \frac{\pi }{2 N} & \frac{8 \pi }{17 N} & \frac{4 \pi }{9 N} & \frac{8 \pi }{19 N} & \frac{2 \pi }{5 N} & \frac{8 \pi }{89 N} \\
 \frac{8 \pi }{77 N} & \frac{\pi }{10 N} & \frac{8 \pi }{83 N} & \frac{4 \pi }{43 N} & \frac{8 \pi }{89 N} & \frac{2 \pi }{23 N} \\
\end{array}
\right)+ O(1/N^2)\;,
\ee
where the rows and columns refer to the values of ${N_\psi, N_\chi}=0,1,\ldots, 5$.  The positive value of the 
possible fixed-point coupling $\alpha_{N_\psi, N_\chi}$  might indicate the
 presence of an IR  CFT.      This table might suggest that the   e models towards the boundaries 
in Fig.~\ref{theories}   are  in the IR  interacting CFT.  
The result    in the  Table below are for type (III) 
\be
\alpha_{N_\psi, N_{\tilde \chi}}=
\left(
\begin{array}{cccccc}
 -\frac{22 \pi }{17  N} & -\frac{24 \pi }{13  N} & -\frac{28 \pi }{5  N} & \frac{40 \pi }{19  N} & \frac{\pi }{2  N} & \frac{8 \pi }{77  N} \\
 -\frac{24 \pi }{13  N} & -\frac{28 \pi }{5  N} & \frac{40 \pi }{19  N} & \frac{\pi }{2  N} & \frac{8 \pi }{77  N} &  \\
 -\frac{28 \pi }{5  N} & \frac{40 \pi }{19  N} & \frac{\pi }{2  N} & \frac{8 \pi }{77  N} &  &  \\
 \frac{40 \pi }{19  N} & \frac{\pi }{2  N} & \frac{8 \pi }{77  N} &  &  &  \\
 \frac{\pi }{2  N} & \frac{8 \pi }{77  N} &  &  &  &  \\
 \frac{8 \pi }{77  N} &  &  &  &  &  \\
\end{array}
\right)+ O(1/N^2)\;.
\ee
  The first row (column) corresponds to $N_\psi=0$ ($N_{\tilde \chi}=0$), so coincides with the first row (column) of the previous table.

Unfortunately, all these putative fix-points (if they really exist) are non-perturbative,  even if $\alpha \ll 1$. The reason is that the coefficients $b_i$ of the beta function scales as $O(N^{i+1})$ so for $\alpha \sim O(1/N)$ all the terms of 
	\be
	\beta(g)=-\frac{g}{12\pi}\left(b_0 \frac{\alpha}{4\pi} + b_1  \left(\frac{\alpha}{4\pi}\right)^2+...\right) \label{eq:beta_function}
	\ee	
	are of the same order. Another, equivalent, way to state this state of matter  is that if \footnote{By adding vector-like matter, e.g. by cranking up $p$ in (\ref{psietabaretadef}) or in (\ref{def:GG}), one can reach a point where $b_0\sim O(1)$, while $b_1$ is still $O(N^2)$. In this case, one obtains $\alpha \sim O(1/N^2)$, and the fix-point is perturbative. The scenario should be similar to the Banks-Zaks fix-point in QCD. We postpone any further discussion on this point to future work.} $b_0 \sim O(N)$, the correct perturbative expansion for large N is the 't Hooft coupling $\lambda = N g^2$ which is  $O(1)$ in these models. We shall not pursue further this problem in this work.

\section{Hypothetical confining, symmetric phase  \label{Hypo}}

{ 
	
Most of the theories studied here  contain a large, non-Abelian flavor symmetry group. Confinement without symmetry breaking, with no condensate formation,  would require  that the spectrum of massless gauge-invariant composite fermions be such that it  matches all the 't Hooft anomaly triangles  of the UV theory. This represents quite a nontrivial  constraint on the IR theory,  and it is somewhat surprising that some models in the family, namely the BY  \cite{Bars:1981se} and the generalized GG  models, apparently allow  for solutions to these constraints, with a simple set of 
massless baryons \cite{Bars:1981se,Appelquist:2000qg}.

 We review first these solutions, and then explain why one cannot expect any solution of this sort  in  other more general models, at least for large N.

Let us start with the $\psi\eta$ model.  The matter content of the theory 
\beq
\psi^{\{ij\}}\,, \quad    \eta_i^A\, , \qquad    i,j = 1,2,\ldots, N\;,\quad A =1,2,\ldots , N+4\;, \label{psietadef}
\eeq
($i, j$ are $SU(N)$ color indices, $A$ is an $SU(8)$ flavor index) transforms as in (\ref{psieta})  under \footnote{The group  (\ref{psietagroup})  is actually not the true symmetry group of our system, but its covering group.  
	Its global structures however contain some redundancies, which must be modded out appropriately in order to eliminate the double counting. 
	They   depend   on whether $N$ is odd or even.  Such an observation was fundamental in the study of the mixed anomalies and generalized anomaly-matching 
	requirement studied in \cite{BKL2}.  
	  }
\be  G_F=SU(N+4) \times  U(1)_{\psi\eta}\;,   \label{psietagroup}
\ee
where $U(1)_{\psi\eta}$ is the anomaly-free combination of  $U(1)_{\psi}$ and $U(1)_{\eta}$,
\be
U(1)_{\psi\eta} : \ \psi\to \rme^{\im (N+4)\alpha}\psi\;, \qquad  \eta \to \rme^{-\im (N+2)\alpha}\eta\;. \qquad \alpha \in \mathbbm{R}
\label{upe0} \;.
\ee  

The surprising feature of this theory is that all the $SU(N+4) \times U(1)_{\psi\eta}$ anomaly triangles  are saturated by a massless "baryon"
\be     {\cal B}^{[AB]}=    \psi^{ij}  \eta_i^A  \eta_j^B \;,\qquad  A,B=1,2, \ldots, N+4\;,\label{baryons101}
\ee
antisymmetric in  $A \leftrightarrow B$, the $SU(N+4)$ indices. The fact that the matching holds, for now, has no simple explanation: the best we can do is check it by inspection of Table~\ref{Simplest0}. 
\begin{table}[h!t]
	\centering 
	\begin{tabular}{|c|c|c |c|c|  }
		\hline
		$ \phantom{{{   {  {\yng(1)}}}}}\!  \! \! \! \! \!\!\!$   & fields  &  $SU(N)_{\rm c}  $    &  $ SU(N+4)$     &   $ {U}(1)_{\psi\eta}   $  \\
		\hline 
		\phantom{\huge i}$ \! \!\!\!\!$  {\rm UV}&  $\psi$   &   $ { \yng(2)} $  &    $  \frac{N(N+1)}{2} \cdot (\cdot) $    & $   N+4$    \\
		& $ \eta^{A}$      &   $  (N+4)  \cdot   {\bar  {\yng(1)}}   $     & $N \, \cdot  \, {\yng(1)}  $     &   $  - (N+2) $ \\
		\hline     
		$ \phantom{ {\bar{   { {\yng(1,1)}}}}}\!  \! \! \! \! \!\!\!$  {\rm IR}&    $ {\cal B}^{[AB]}$      &  $  \frac{(N+4)(N+3)}{2} \cdot ( \cdot )    $         &  $ {\yng(1,1)}$        &    $ -N    $   \\
		\hline
	\end{tabular}
	\caption{\footnotesize  Chirally symmetric phase of the  $(1,0)$  model.   
		The multiplicity, charges and the representation are shown for each  set of fermions. $(\cdot)$ stands for a singlet representation.
	}\label{Simplest0}
\end{table}

It is possible to generalize this solution for all the  BY models $(N_{\psi}=1, ... )$, defined in (\ref{psietabaretadef}).
The fermions
\beq
 \psi^{[ij]}\,, \quad \eta_i^A \quad  \xi^{i,b}\, , \qquad    i,j = 1,2,\ldots, N\;,\quad A=1, 2,\ldots, N+p+4\;\quad b =1,2,\ldots , p\;,
\eeq
transform under an enlarged global symmetry 
\be
G_F=\begin{cases} SU(N+5)_{\eta}  \times   U(1)_{\psi\eta}\times  U(1)_{\psi\xi}\;, \quad \text{for}\;\; p=1\;, \\
	SU(N+4+p)_{\eta}  \times  SU(p)_{\xi}  \times  U(1)_{\psi\eta}\times  U(1)_{\psi\xi}\;,\quad \text{for}\;\; p>1\;,  
	 \label{grouplocal}
	\end{cases}
\ee
Here, again,  
\be
U(1)_{\psi\eta} : \qquad  \psi\to \rme^{\im (N+4+p)\alpha}\psi\;, \quad  \eta \to \rme^{-\im (N+2)\alpha}\eta\;, \label{upe}
\ee  
 with $\alpha \in \mathbbm{R}$, and 
    \be
U(1)_{\psi\xi} : \qquad \psi\to \rme^{\im p \beta}\psi\;, \quad  \xi \to \rme^{-\im (N+2)\beta}\xi\;, \label{upx}
\ee  
with $\beta \in \mathbbm{R}$, is a particular choice of the two, Abelian, anomaly free symmetries.

In this case the solution \ref{baryons101} generalizes to
\be    
{({\cal B}_{1})}^{[AB]}=    \psi^{ij}   \eta_i^{A}  \eta_j^{B}\;,
\qquad {({\cal B}_{2})}^{a}_{A}=    \bar{\psi}_{ij}  \bar{\eta}^{i}_{A}  \xi^{j,a} \;,
\qquad {({\cal B}_{3})}_{\{ab\}}=    \psi^{ij}  \bxi_{i,a}  \bxi_{j,b}  \;,
\label{baryons10}
\ee
where the first baryon is anti-symmetric in $A \leftrightarrow B$ and the third is  symmetric in $a \leftrightarrow b$. Their charges are given in Table  \ref{sir}, and again, one must check all the 't Hooft anomaly matching condition by inspection.

\begin{table}[h!t]
	\centering 
	\small{\begin{tabular}{|c|c|c |c|c| c|c| }
			\hline
			\su      &  $SU(N)_{\rm c}  $    &  $ SU(N+4+p)$    &  $ SU(p)$     &   $ {U}(1)_{\psi\eta}   $  &   $ {U}(1)_{\psi\xi}   $  \\
			\hline     
			\sbuu     $ {{\cal B}_{1}}$      &  $   \frac{(N+4+p)(N+3+p)}{2} \cdot ( \cdot )    $         &  $ {\yng(1,1)}$        &    $  \frac{(N+4+p)(N+3+p)}{2} \cdot ( \cdot )     $   &  $-N+p $ & $p$ \\
			\hline     
			\sbbu   $ {{\cal B}_{2}}$   &    $ (N+4+p) p\cdot ( \cdot )$     &       $  p\cdot   \bar{\yng(1)}$   &     $ (N+4+p)  \cdot {\yng(1)}$      & $-(p+2)$ & $-(N+p+2)$ \\
			\hline     
			\sbbu  ${{\cal B}_{3}}$     &  $ \frac{p (p+1)}{2}  \cdot ( \cdot )   $    &    $ \frac{p (p+1)}{2}  \cdot ( \cdot )   $       &    $ \bar{\yng(2)}$       & $N+4+p$ & $2N +4 + p$\\
			\hline
	\end{tabular}}
	\caption{\footnotesize  Chirally symmetric phase of the  BY  model. 
	}\label{sir}
\end{table}
All anomaly triangles  are indeed saturated  by these candidate massless composite fermions,  as  seen in Table~\ref{suvsir}, taken from \cite{BKL2}.

There is a similar solution also for all the generalized  Georgi-Glashow models \ref{def:GG}.  
\begin{table}[h!t]
	\centering
	\tiny{\begin{tabular}{|c|c|c|}
			\hline
			\su      &  UV    &  IR  \\
			\hline 
			\su $ SU(N+4+p)^3$     &   $ N $     &  $ N+p -p $ \\
			\su $ SU(p)^3$     &   $ N $     &  $ N+4+p -(p+4) $ \\
			\su $ SU(N+4+p)^2 - {U}(1)_{\psi\eta}$     &   $ - N(N+2)  $     &  $ -(N+2+p)(N-p) -p(p+2) $ \\
			\su $ SU(N+4+p)^2- {U}(1)_{\psi\xi}$     &    $ 0 $     &  $  (N+2+p)p -p(N+p+2)  $ \\
			\su $ SU(p)^2- {U}(1)_{\psi\eta}$     &   $ 0 $     &  $ -(N+4+p)(p+2) +(p+2)(N+p+4)  $ \\
			\su $ SU(p)^2- {U}(1)_{\psi\xi}$     &   $ - N(N+2)   $     &  $   -(N+4+p)(N+p+2) +(p+2)(2N+p+4)$ \\
			\su $ {U}(1)_{\psi\eta}^3$     &   $ \frac{N(N+1)}{2}(N+4+p)^3 - N(N+4+p) (N+2)^3 $     &  $ - \frac{(N+4+p)(N+3+p)}{2} (N-p)^3 -(N+4+p) p (p+2)^3 + $ \\\
			\su      &    
			&  $ +  \frac{p (p+1)}{2}(N+4+p)^3$ \\
			\su $ {U}(1)_{\psi\xi}^3$     &   $ \frac{N(N+1)}{2}p^3 - N p (N+2)^3  $     &  $    \frac{(N+4+p)(N+3+p)}{2} p^3 -(N+4+p) p (N+p+2)^3  +$ \\
			\su    &     
			&  $    \  +  \frac{p (p+1)}{2}(2N+4+p)^3$ \\
			\su $ {\rm Grav}^2-{U}(1)_{\psi\eta} $     &  $ \frac{N(N+1)}{2}(N+4+p)  - N(N+4+p) (N+2)  $     &  $ - \frac{(N+4+p)(N+3+p)}{2} (N-p)  -(N+4+p) p (p+2) + $ \\
			\su      &   
			&  $ +  \frac{p (p+1)}{2}(N+4+p) $ \\
			\su $ {\rm Grav}^2-{U}(1)_{\psi\xi}$       &   $\frac{N(N+1)}{2}p - N p (N+2)  $     &  $    \frac{(N+4+p)(N+3+p)}{2} p -(N+4+p) p (N+p+2)  +$ \\
			\su    &     
			&  $    \  +  \frac{p (p+1)}{2}(2N+4+p)$ \\
			\su $ SU(N+4+p)^2-({\mathbbm Z}_{N+2})_{\psi}  $     &  $0$    & $N+2+p-p = 0 \ {\rm mod} \ N+2$    \\
			\su $ SU(p)^2- ({\mathbbm Z}_{N+2})_{\psi} $     &   $0$    & $-(N+4+p) + p+2  = 0 \ {\rm mod} \ N+2 $     \\
			\su $ {\rm Grav}^2-({\mathbbm Z}_{N+2})_{\psi}  $     &  $1$      & $1-1+1$  \\
			\hline   
	\end{tabular}}
	\caption{\footnotesize  Anomaly matching checks for the IR chiral symmetric phase of the BY  model.    
	}\label{suvsir}
\end{table}

The problem  is that we do not have any physical understanding (apart from that of the general principle) of why and how these matching equations are satisfied.
This in in stark contrast to the case of the dynamical Higgs phase discussed below, Sec.~\ref{DHiggs}.

Moreover, it seems that these solutions \textit{cannot} be extended to other models. One can give a simple argument for this, for large N.
}

Let us  first  discuss the possibility of  an IR phase of this sort  for the $\psi\chi\eta$ model, that is the $(N_{\psi},N_{\chi})=(1,1)$ model. 
Take for example the ‘t Hooft $[SU(8)_{\eta}]^3$ anomaly, where $SU(8)_{\eta}$ is the global symmetry acting on the anti-fundamentals $\eta$. This anomaly  is $N$ in UV, because they are anti-fundamentals fields under $SU(N)$. To saturate it in the IR it would require $\propto N$ distinct gauge invariant composite fermions charged under $SU(8)_{\eta}$.   
 The possibility  that the system confines, with no global symmetry breaking and with some massless ``baryons"  saturating the 't Hooft anomalies, thus does not appear to be
 likely here, at least in the large $N$ limit  \cite{Goity:1985tf,Eichten:1985fs,BK}.

Next, consider   the possibility of  such IR phase  for the $(N_{\psi}>1,0)$ model, that is $N_{\psi}$ families of the $\psi\eta$ multiplet. 
We do have simple gauge invariant fermion, e.g. $\psi\eta\eta$  in    $\tiny{\yng(2)}$    or  $\tiny{\yng(1,1)}$       of $SU(N_{\psi}(N+4))_{\eta}$ and in  fundamental of     of  $SU(N_{\psi})_{\psi}$, we also have composite of these ones, and more made with the epsilon tensor.
 Take for example the $[SU((N_{\psi}(N+4))_{\eta}]^3$ anomaly which is $N$ in the UV. For this anomaly    $\psi\eta\eta$  would give a contribution $(N_{\psi}((N_{\psi}(N+4) \mp 4)$ in the IR.  Clearly  $\psi\eta\eta$, no matter its representation, cannot saturate these anomalies (only in the special case s $N_{\psi}=1$ of the $\psi\eta$ model the saturation works). 
Take for example the ‘t Hooft $[SU((N_{\psi} )_{\psi}]^3$  anomaly which is $N(N+1)/2$ n the UV,   $\psi\eta\eta$ would give a contribution $(N_{\psi}(N+4)((N_{\psi}(N+4) \mp 1) /2$  in the IR.
Cannot be excluded, but confinement without symmetry breaking for the $(N_{\psi}>1,0)$ model is not a plausible solution for large $N$.
Similar considerations for the $(N_{\chi}>1,0)$ model, that is $N_{\chi}$ families of the $\chi \tilde{\eta}$ multiplet, or any model where we have more than one $\psi$ or $\chi$, essentially any model of the type (\ref{general1}), (\ref{general2}) apart from BY and GG for which we saw the easy solution before.

Any solution, even if any can be found, 
 is rather unnatural, very contrived and highly dependent of $N$. See for example \cite{Bai:2021tgl} with three copies of $\chi \eta$ model with $N=5$, this is the  Georgi-Glashow GUT model with three families, which has a clear interest for phenomenological purposes.

Most significantly,   the recent series of studies based on the generalized 't Hooft  (mixed-anomaly) matching  conditions, see \cite{BKL2,BKL4,BKL5,BKLReview},
cast serious doubt  on the consistency of such a confining, fully symmetric phase, reviewed in this section.

\section{Dynamical Higgs phase   \label{DHiggs}  }

{
	
A natural candidate for the IR dynamics of the  BY and GG  models  is what might be called dynamical Higgs phase.  
Namely,  as the interactions become strong towards the infrared, bi-fermion condensates are formed so as to break both color $SU(N)$ and (part of) the global symmetry of the system.  A particularly interesting possibility is the  color-flavor locked  bi-fermion condensates are formed and characterize the IR system, as studied in  \cite{BK,BKL2,BKL4,BKL5,BKLReview}.

A remarkable property of this dynamical Higgs phase  
is  that all the 't Hooft anomaly matching equations are automatically satisfied.  The 
 IR spectrum of the massless fermions  in such a system  is found  by decomposing the UV fermion representations as  direct  sums of the  irreps of the unbroken symmetry group (gauge and flavor).  All Dirac-like pairs of fermions with respect to the unbroken group can be assumed to get mass and decouple.  The spectrum of the remaining fermions  is then seen to be  {\it  identical}   in  UV and IR (their charges and multiplicities) 
hence  the anomaly matching  (including the generalized, mixed anomalies) is automatic, and does not require checking any arithmetic equations such as in Table~\ref{suvsir}.
In other words, we understand why the 't Hooft anomaly constraints are satisfied. 
The matching of the anomalies of the spontaneously broken part of the symmetries, instead, is taken care  of   by the coupling of the Nambu Goldstone bosons with the background gauge fields.

As the 't Hooft anomaly matching requirement does not lead to  any new constraint here, it is difficult to know  which  particular condensate and which symmetry-breaking pattern is actually  realized. To determine the correct one, one can appeal to   several heuristic arguments:
\begin{itemize}
	
	\item The Maximally Attractive Channel (MAC) criterion \cite{Raby:1979my} suggests that the condensate that actually forms is the one that maximizes (in the absolute value)  the quantity
	\be
	C_2(\mathcal R_c)-C_2(\mathcal R_1)-C_2(\mathcal R_2)   \label{force}
	\ee
	where $C_2(\mathcal R)$ is the quadratic Casimir of the irrep $\mathcal R$.  
	It represents the strength of the one-gluon 
 exchange force  (\ref{force})  in various bifermion  (made of $(\mathcal R_1)$ and $(\mathcal R_2)$)  composite-scalar   $(\mathcal R_c)$  channels.
	
Just to have some quantitative idea, let us
compare the strength of the attraction   (\ref{force})  in various bifermion scalar  channels, formed by two out of the three types of fermions, $\psi$, $\chi$ and $\eta$. Some of the most probable
 channels are  
  \bea & & A:  \qquad  \psi \left(\raisebox{-2pt}{\yng(2)}\right) \, \psi \left(\raisebox{-2pt}{\yng(2)}\right)   \quad  {\rm forming}  \quad   \,  \raisebox{-6pt}{\yng(2,2)}\;;
\nonumber \\  & & B:  \qquad  \chi \left(\bar{\raisebox{-9pt}{\yng(1,1)}}\right) \, \chi  \left(\bar{\raisebox{-9pt}{\yng(1,1)}}\right)  \qquad \ \   {\rm forming}  \quad \bar{\raisebox{-12pt}{\yng(1,1,1,1)}}\;;
\nonumber\\  &&C:  \qquad  \eta \left(\bar{\raisebox{-2pt}{\yng(1)}}\right) \, \eta \left(\bar{\raisebox{-2pt}{\yng(1)}}\right)   \quad  \qquad \ \  {\rm forming}  \quad    \bar{\raisebox{-9pt}{\yng(1,1)}} \;;
\nonumber\\ && D:  \qquad  \chi  \left(\bar{\raisebox{-9pt}{\yng(1,1)}}\right) \, \eta\left(\bar{\raisebox{-2pt}{\yng(1)}}\right)   \qquad  \  \  {\rm forming}  \quad   \bar{\raisebox{-9pt}{\yng(1,1,1)}}\;;
\nonumber \\  && E:  \qquad   \psi  \left(\raisebox{-2pt}{\yng(2)}\right)   \, \chi \left(\bar{\raisebox{-9pt}{\yng(1,1)}}\right)  \quad \  \ {\rm forming ~ an ~ adjoint ~ representation}\,\, ({\tilde \phi})\;;  
\nonumber\\ && F:  \qquad \psi  \left(\raisebox{-2pt}{\yng(2)}\right)   \, \eta \left(\bar{\raisebox{-2pt}{\yng(1)}}\right)  \quad   \quad\  {\rm forming}  \quad   \raisebox{-2pt}{\yng(1)}\; \,\, \,  ({\phi});
\nonumber\\ && G:  \qquad  \tilde{\eta}  \left(\raisebox{-2pt}{\yng(1)}\right)   \, \eta  \left(\bar{\raisebox{-2pt}{\yng(1)}}\right)  \quad   \quad  {\rm forming}  \quad   
{(\cdot)}\;
 \,\, \,  ({\rm singlet})\;.   
\label{MAC1}
 \eea
The one-gluon exchange strength is, in the six cases above, proportional to  
 {\footnotesize  
 \bea  & &  A:  \qquad     \frac{2 (N^2-4)}{N} -     \frac{ (N+2)(N-1)}{N}  -     \frac{ (N+2)(N-1)}{N}    = - \frac{2 (N+2)}{N}  \;;
\nonumber \\ & &  B:  \qquad     \frac{2 (N+1)(N-4))}{N} -     \frac{ (N+1)(N-2)}{N}  -     \frac{ (N+1)(N-2)}{N}    = - \frac{4 (N+1)}{N}  \;;
\nonumber\\   & &C:  \qquad     \frac{(N+1)(N-2))}{N} -     \frac{ N^2-1}{2N}  -     \frac{ N^2-1}{2N}    = - \frac{ N+1}{N}  \;;
\nonumber\\  & & D:  \qquad     \frac{3 (N+1)(N-3))}{2N} -     \frac{ N^2-1}{2N}  -     \frac{ (N+1)(N-2)}{N}    = - \frac{2N+2}{N}  \;;
\nonumber\\  & &  E:  \qquad   N-     \frac{ (N+2)(N-1)}{N}  -     \frac{ (N+1)(N-2)}{N}    =   - \frac{N^2-4}{N}  \;;
\nonumber \\ & &  F:  \qquad     \frac{N^2-1}{2N} -     \frac{N^2-1}{2 N}  -     \frac{ (N+2)(N-1)}{N}    = - \frac{(N+2)(N-1)}{N}  \;;
\nonumber \\ & &  G:  \qquad    0 -    \frac{N^2-1}{2N} -     \frac{N^2-1}{2 N}   =  -     \frac{N^2-1}{ N}    \;,  \label{MAC2}
 \eea
 }
 respectively.
 We note that the      ${\tilde \phi}$   ($\psi  \chi$)    and  $\phi$   ($\psi \eta$) condensates considered by us   (cases E and F, respectively), together with 
 the ``quark-antiquark condensate"  (case G),      correspond  precisely to the three most attractive channels, at large $N$. 
 Their attraction strength  scales as  $O(N)$ in contrast to the other four channels which scale as   $O(1)$.

	Note in particular that the bifermion  $\psi\eta$ (or $\chi\eta$) condensate channels  assumed  in the $\psi\eta$ (and $\chi\eta$) models, see (\ref{cflocking}), (\ref{cflocking2})  below,  have the same attraction  strength 
	as  in  the familiar color-singlet  $\brc {\bar q} q \ckt$  condensates in the standard  QCD.

	\item A second prescription is that the right condensate to consider is the one that minimizes $a_{IR}$, a quantity defined at a fixed point (either IR or UV) from the conformal anomaly. In a weakly coupled theory, $a$ can be computed from the massless spectrum: 
	\be
	a=(\#\text{ bosons}) + \frac{11}{2}(\# \text{Weyl fermions})+62 (\# \text{vector bosons})\;.
	\ee
	This prescription arises from the $a$-theorem, proposed a long time ago \cite{Cardy:1988cwa} and proven only recently \cite{Komargodski:2011vj}. Unfortunately, the theorem imposes only that $a_{UV}>a_{IR}$, but this does not mean that  $a_{IR}$ is the smallest possible between a set of candidates: it should simply be smaller than the $a$ in the UV.

	\item A related idea is the $ACS$ prescription \cite{Appelquist:1999vs}. 

\end{itemize}

}

The simplest example of color-flavor locking phases is the $\psi\eta$ model. It is natural to assume  (see (\ref{MAC1}), (\ref{MAC2}))  that
a bifermion condensate 
\be    \brc  \psi^{\{ij\}}   \eta_i^b \ckt =\,   c \,  \Lambda^3   \delta^{j b}\ne 0\;,   \qquad   j, b =1,2,\dots  N\;,   \qquad c \sim O(1) \label{cflocking}
\ee
forms, breaking  the gauge-flavor symmetries as 
\beq
SU(N)_c \times SU(N+4)\times U(1)_{\psi\eta} \rightarrow
SU(N)_{\rm cf} \times  SU(4)_{\rm f}  \times U(1)^{\prime} \,.     \label{hs}
\eeq   
Here $U(1)^{\prime}$ symmetry is the diagonal combination of $U(1)_{\psi \eta}$ and the elements of $SU(N+4)$ generated by 
\be  \left(\begin{array}{cc}- 2\,   \mathbf{1}_N &  \\ & \frac N2 \mathbf{1}_4\end{array}\right) \;
\ee
left unbroken by the condensate.

In the IR, alongside the  $8N+1$  Nambu-Goldstone (NG) bosons associated with the breaking \ref{hs}, there are a few  leftover massless fermions that enforce  the 't Hooft anomaly matching of the unbroken symmetry. As explained already, the decomposition of the UV fermions as a direct sum of the irreps of the unbroken group, see the upper  half of Table \ref{SimplestBis},  is sufficient to show that  all the matching requirements are met by construction. 
\begin{table}[h!t]
	{
		\centering 
		\begin{tabular}{|c|c|c |c|c| }
			\hline
			$ \phantom{{{   {  {\yng(1)}}}}}\!  \! \! \! \! \!\!\!$   & fields   &  $SU(N)_{\rm cf}  $    &  $ SU(4)_{\rm f}$     &   $  U(1)^{\prime}   $     \\
			\hline
			\phantom{\huge i}$ \! \!\!\!\!$  {\rm UV}&  $\psi$   &   $ { \yng(2)} $  &    $  \frac{N(N+1)}{2} \cdot   (\cdot) $    & $   N+4  $     \\
			& $ \eta^{A_1}$      &   $  {\bar  {\yng(2)}} \oplus {\bar  {\yng(1,1)}}  $     & $N^2 \, \cdot  \, (\cdot )  $     &   $ - (N+4) $      \\
			&  $ \eta^{A_2}$      &   $ 4  \cdot   {\bar  {\yng(1)}}   $     & $N \, \cdot  \, {\yng(1)}  $     &   $ - \frac{N+4}{2}  $    \\
			\hline 
			$ \phantom{{\bar{ \bar  {\bar  {\yng(1,1)}}}}}\!  \! \!\! \! \!  \!\!\!$  {\rm IR}&      $ {\cal B}^{[A_1  B_1]}$      &  $ {\bar  {\yng(1,1)}}   $         &  $  \frac{N(N-1)}{2} \cdot  (\cdot) $        &    $   -(N+4) $      \\
			&   $ {\cal B}^{[A_1 B_2]}$      &  $   4 \cdot {\bar  {\yng(1)}}   $         &  $N \, \cdot  \, {\yng(1)}  $        &    $ - \frac{N+4}{2}$      \\
			\hline
		\end{tabular}  
		\caption{\footnotesize   Color-flavor locked phase in the $(1,0)$ model.
			$A_1$ or $B_1$  stand for  $1,2,\ldots, N$,   $A_2$ or $B_2$ the rest of the flavor indices, $N+1, \ldots, N+4$.   
		}\label{SimplestBis}
	}
\end{table}

Generically, the symmetric part of  $\eta^{A_1}$ and $\psi$ form a massive Dirac pair,   leaving in IR the symmetric part of $\eta^{A_1}$ and $\eta^{A_2}$. 
At this point one can always dress the IR particles with the condensate to make them invariant under the UV gauge group: 
\be
\mathcal{B}^{[A_1, B_1]}=\psi^{ij}\eta_j^{[A_1}\eta_i^{B_1]} \quad \text{and} \quad \mathcal{B}^{[A_1, B_1]}=\psi^{ij}\eta_j^{[A_1}\eta_i^{B_2]}
\ee 
These fermions can be thought of as what remains of the massless baryon \ref{baryons101}, if one enforces the symmetry breaking \ref{hs}. However we should stress that these two phases are different: the symmetries and the massless spectrum distinguish them.

{
	One can consider a similar solution for the $\tilde \chi \eta$ model. Again, one has a bi-fermion condensate
	\be    \brc  \tilde\chi^{[ij]}   \eta_i^b \ckt =\,   c \,  \Lambda^3   \delta^{j b}\ne 0\;,   \qquad   j, b =1,2,\dots  N\;,   \qquad c \sim O(1) \label{cflocking2}
	\ee
	that breaks
	\be
	SU(N)_c \times SU(N-4)_\eta \times U(1)_{\chi\eta} \rightarrow SU(4)_c \times SU(N-4)_{cf} \times \tilde U(1)  
	\ee
	where $\tilde U(1)$ is a particular combination of $U(1)_{\chi\eta}$ and a $SU(N)_c$ rotation generated by
	\be
	\tilde Q=  Q_{\chi\eta}+  \frac{1}{2}\left(\begin{array}{cc} 4\,\cdot   \mathbf{1}_{(N-4) \times (N-4)} &  \\ & -(N-4)\;\cdot  \mathbf{1}_{4 \times 4} \end{array}\right)\;. 
	\ee
	
	Notice that the structure of the global symmetry group is the same as we have in UV: there are no NGBs and the charges of any gauge invariant operator under $SU(N-4)_{cf}\times \tilde U(1)$ and under $SU(N-4)\times U(1)_{\chi\eta}$ are identical\footnote{Up to a complex conjugation. This is because we defined the generator of $SU(N-4)_{cf}$ as $T^a_{cf}=T^a_c - T^a_{f}$.}.  
	
	To figure out the spectrum it is useful to decompose the fundamental fermions under the leftover symmetry group, as in Table \ref{tab:chietahiggs}. One can see that $\chi_1$ and $\eta_2$, $\chi_2$, and $\eta_3$ form complex conjugate pairs, so they can be gapped together. Similarly, $\eta_3$ is self-conjugate, so, as $SU(4)$ confines, also $\chi_3$ becomes massive, and disappears from the spectrum. In the IR only $\eta_1$ remains.

	\begin{table}[h!t]
		{
			\centering 
			\begin{tabular}{|c|c|c |c|c|  }
				\hline
				$ \phantom{{{   {  {\yng(1)}}}}}\!  \! \! \! \! \!\!\!$   & fields   &  $SU(4)_{c}$    &  $SU(N-4)_{\rm cf}$     &   $\tilde U(1)$    \\
				\hline
				\phantom{\huge i}$ \! \!\!\!\!$  {\rm UV}&  $\chi_1$ & $(\cdot)$ & $\yng(1,1)$ \phantom{$\bar{\yng(1,1)}$}$ \!\!\! \!\!\!\!\!\!$ & $N$\\
				& $\chi_2$ & $\yng(1)$   & $\yng(1)$   & $N/2$ \\
				& $\chi_3$ & $\yng(1,1)$ & $(\cdot)$   & $0$ \\
				& $\eta_1$ & $(\cdot)$   & $\bar{\yng(2)}$ & $-N$ \\
				& $\eta_2$ & $(\cdot)$   & $\bar{\yng(1,1)}$   & $-N$ \\
				& $\eta_3$ & $\bar{\yng(1)}$   & $\bar{\yng(1)}$   & $-N/2$\\
				\hline
			\end{tabular}  
			\caption{\footnotesize   Color-flavor locked phase in the $(0,1)$ model.
				$A_1$ or $B_1$  stand for  $1,2,\ldots, N$,   $A_2$ or $B_2$ the rest of the flavor indices, $N+1, \ldots, N+4$.  
			}\label{tab:chietahiggs}
		}
	\end{table}

	Dressing $\eta_1$ with the condensate 
	\be
	\left(\chi^{[ij]}\eta_j^{\{A}\right) \eta_i^{B\}}=\mathcal{B}^{\{AB\}}
	\ee
	would make it gauge invariant.  Such  ``baryons"   are identical to the massless composite fermions which would saturate the full 't Hooft anomaly triangles 
	in the hypothetical, confining flavor-symmetric vacuum, as those discussed in Sec.~\ref{Hypo},  for the ${\tilde{\chi}} \eta$ model  (see \cite{BKL2,BKL5} for more details).  
		This might lead some to suspect that in the ${\tilde \chi} \eta$  model  the so-called complementarity \cite{FS} is at work, i.e., that dynamical Higgs and confinement without symmetry breaking might be actually the same phase. 
	
	As discussed carefully in  Section 4 of \cite{BKL5}, however, there are reasons to believe that these two phases  are
	actually physically distinct.    There is no complementarity in the  ${\tilde \chi} \eta$ model.   Only one of them can be the correct IR phase of the theory. 
	 As discussed in \cite{BKL2,BKL4,BKL5}  the dynamical Higgs phase appears to describe correctly the IR physics of the  ${\tilde \chi} \eta$ model.

}

\section{Dynamical Abelianization   \label{DA} }

Another remarkable infrared phase which might describe the physics of some of the models in the infrared,  is dynamical Abelianization,  a phenomenon 
familiar from the exactly solved ${\cal N}=2$   supersymmetric gauge theories \cite{SW1,SW2}.   What  is perhaps not quite familiar is the fact that 
dynamical Abelianization can occur in  some class of  nonsupersymmetric, chiral gauge theories  we are interested in.

The first carefully studied case concerns the   $\psi\chi\eta$ model, (\ref{psichieta}).   We assume that  
a  bifermion condensate in the adjoint representation forms: 
\be   
\langle  \psi^{ik} \chi_{kj}   \rangle  = \Lambda^3   \left(\begin{array}{ccc}
      c_1  &  &  \\     & \ddots   &   \\   &  &    c_{N}
  \end{array}\right)^i_j   \;, \qquad  \brc  \psi^{ij}  \eta_j^A  \ckt    = 0
  \;,   \label{psichicond}  
\ee
\be         c_n    \in {\mathbbm C}\;,     \qquad     \sum_n c_n =0\;, \qquad       c_m - c_n  \ne 0\;, \ \     m \ne n  \;,  \label{psichicondBis}  
\ee
(with no other particular  relations among $c_j$'s), inducing dynamical Abelianization of the system. 
 The unbroken gauge group $U(1)^{N-1}$   is generated by the Cartan subalgebra.  
More precisely \cite{BKLDA},
we require that  the condensate (\ref{psichicond}),  (\ref{psichicondBis})  induce  the symmetry breaking  
\be    
SU(N)    \to    U(1)^{N-1} \;.      \label{SUNbreaking}    \ee
As the effective composite scalar fields  $\phi \sim  \psi\chi$  are in the adjoint representation of $SU(N)$,   it can be parametrized  as a linear combination, 
\be  \phi \sim  \psi\chi  =    \phi^A  T^A  =   \phi^{(\alpha)} E_{\alpha} +     \phi^{(-\alpha)} E_{-\alpha} +     \phi^{(i)} H^i \;,\label{last}  
\ee
where  $\phi^A$  are complex  fields and $T^A$ are  the Hermitian  generators of $SU(N)$ in the fundamental representation ($A= 1,2,\ldots, N^2-1$).
$ E_{\pm\alpha} $ are the raising and lowering generators in the Cartan basis, associated with the various root vectors   $\alpha$.
 The condition for the  dynamical Abelianization (\ref{psichicond})   is    that 
 the fields that condense are in the Cartan subalgebra, 
 \be   \phi  \sim       \psi \chi  =        \phi^{(i)} H^i \;, \qquad     \brc  \phi^i  \ckt  \ne  0\;, \qquad   \forall  i \;, \label{precise1}  
\ee
 whereas   
 \be     \brc  \phi^{(\alpha)}   \ckt =    \brc  \phi^{(-\alpha)}   \ckt =0\;,  \qquad  \forall \alpha\;. \label{precise2}  
 \ee
See \cite{BKLDA}  for more about  the associated (would-be) NG bosons.

\begin{figure}[h!]
\begin{center}
\includegraphics[width=5.8in]{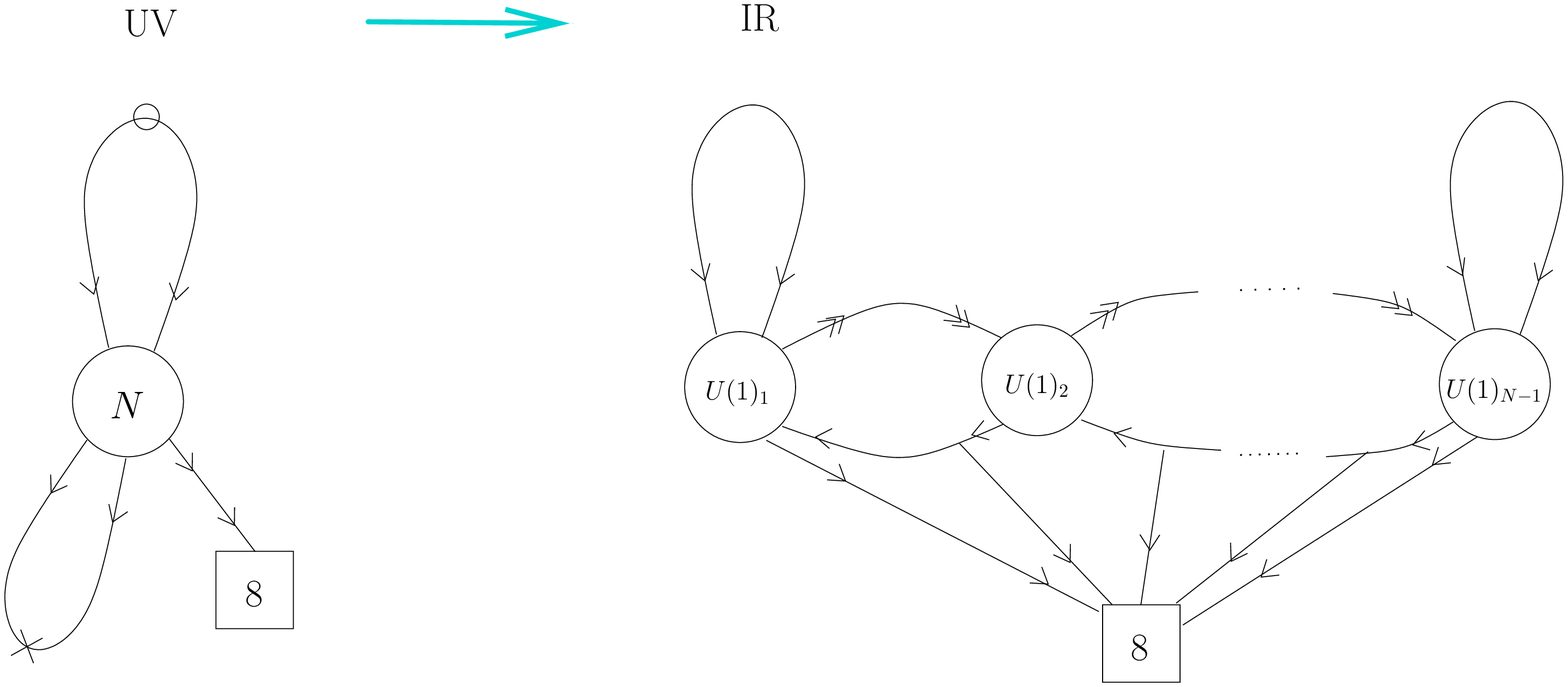}
\caption{\small    $U(1)$'s quiver model representing dynamical Abelianization for the $\psi\chi\eta$ model   }
\label{abquiv}
\end{center}
\end{figure}

The fields  $\eta^A_i$ which do not participate in the condensate  remain  massless and  weakly coupled  to the gauge bosons from the Cartan subalgebra which we will refer to as the photons.   Also, some of the fermions $\psi^{ij}$  do not participate in the condensates. Due to the fact that 
$\psi^{\{ij\}}$ are symmetric whereas $\chi_{[ij]}$ are antisymmetric,  actually only the nondiagonal elements of   $\psi^{\{ij\}}$  condense and get mass. The diagonal 
fields $\psi^{\{ii\}}$, $i=1,2,\ldots, N$, together with  all of  $\eta^a_i$  remain massless.   Also there is one physical  NG boson. 
\begin{table}[h!t]
  \centering 
  \begin{tabular}{|c|c|c |c| }
\hline
$ \phantom{{{   {  {\yng(1)}}}}}\!  \! \! \! \! \!\!\!$   & fields      &  $ SU(8)$     &   $ {\tilde U}(1)   $  \\
 \hline
  \phantom{\huge i}$ \! \!\!\!\!$  {\rm UV}& $\psi$      &    $   \frac{N(N+1)}{2} \cdot  (\cdot) $    & $  \frac{N(N+1)}{2} \cdot (2)$    \\
  &  $\chi$      &    $   \frac{N(N-1)}{2} \cdot  (\cdot) $    & $  \frac{N(N-1)}{2} \cdot (-2)$        \\
 &$ \eta^{A}$      &   $ N \, \cdot  \, {\yng(1)}  $     &   $  8N \cdot  (-1) $ \\
   \hline 
  \phantom{\huge i}$ \! \!\!\!\!$  {\rm IR}&       $ \psi^{ii}  $      &  $ N \cdot ( \cdot)   $        &    $  N \cdot (2) $   \\
     &  $ \eta^{A} $      &  $ N \, \cdot  \, {\yng(1)}  $        &    $  8  N \cdot (-1) $   \\
\hline
\end{tabular}  
  \caption{\footnotesize  Full dynamical Abelianization in the $\psi\chi\eta$ model.}
\label{Simplest}
\end{table}
All of the anomaly triangles,    $[SU(8)]^3$,    $  {\tilde U}(1)-[SU(8)]^2$,   $[{\tilde U}(1)]^3$,      ${\tilde U}(1)-[{\rm gravity}]^2$,   
are  easily seen to match, on inspection of  Table~\ref{Simplest}.  Perhaps the only non-trivial ones are the ones that do not involve $SU(8)$.  
 The unbroken gauge group $ \prod_{\ell=1}^{N-1}  U(1)_{\ell}  $  is generated by the subalgebra,  
  \bea    t^1 =  \frac{1}{2} {\rm diag}\left( 1,  -1 , 0,  \dots ,  0 \right)\;, 
  \quad  t^2 =  \frac{1}{2 }  {\rm diag}\left( 0 , 1 ,-1  ,0,  \dots , 0\right)\;,
   \nn 
  \\  \qquad           \cdots,         \quad  t^{N-1}  =  \frac{1}{2 }  {\rm diag}\left( 0 , \dots   , 0, 1 , -1\right)\;.
  \label{gent}
  \eea  
In this basis  the IR theory can be represented as the $U(1)$ quiver diagram of figure \ref{abquiv}. 

The dynamical Abelianization proposal  (\ref{psichicond}),  has been shown to be fully consistent with a stronger requirement of the matching of the 
mixed anomalies,    in \cite{BKLDA}.   Also, as discussed in \cite{BK,BKL1,Anber:2021iip}  dynamical Abelianizatio may well describe the IR phase of many other models.

\section{Non-Abelian gauge groups in the IR    \label{new} }

Let us now  discuss  generalizations of the dynamical Abelianization  in  $(N_{\psi},N_{\chi}) $ models of type $I$ and $II$. All these models have a $\psi \chi$  composite scalar  in the adjoint representation.  The condensation of such a composte adjoint scalar field can lead to dynamical Abelianization, as discussed in the previous section.   It is, however, possible 
that the condensation of the adjoint scalar  $\brc \psi \chi  \ckt $   leads to more general types of  gauge symmetry breaking. 
Let us now discuss the possibility that  the RG flow actually brings the system  towards more general  low-energy  effective gauge groups, containing various nonAbelian factors.

A type (I) theory has fields
\bea   && \qquad \qquad \qquad \qquad \quad  \psi^{\{ij\},a}\;, \qquad  \chi_{[ij]}^b\;, \qquad    \eta_i^c\;, \nn \\ &&   a =1,\ldots N_{\psi}  \;, \quad \  b=1,\ldots N_{\chi} \;, \quad  \  c =1,\ldots , N_{\psi}(N+4) -N_{\chi}(N-4)  \;.  
\eea
We assume that only one of the $\psi$'s pairs with one of the $\chi$'s   condense. 
The $\psi\chi$ condensate breaks the global symmetry as 
\bea
&& SU(N_{\psi}) \times SU(N_{\chi}) \times SU(N_{\psi}(N+4) -N_{\chi}(N-4) ) \times U(1)^2  \nn  \\
&& \longrightarrow SU(N_{\psi}-1) \times SU(N_{\chi}-1) \times SU(N_{\psi}(N+4) -N_{\chi}(N-4) ) \times U(1)^3 \;.
\eea
We assume that the condensate can be brought in a diagonal form
\be   
\langle  \psi^{ik,1} \chi_{kj}^1   \rangle  = \Lambda^3   \left(\begin{array}{ccc}
      c_1  &  &  \\     & \ddots   &   \\   &  &    c_{N}
  \end{array}\right)^i_j   \,     \ne  0
  \;,   \label{psichicondna}  
\ee
\be         c_n    \in {\mathbbm C}\;,     \qquad     \sum_n c_n =0\;,  
\ee
as in the dynamical-Abelianization case, but this time we allow the possibility for some  degeneracy among  $c$'s. Let us assume  for instance  that there  a block of $n$ coefficients which are degenerate:
\beq
c_1 = c_2 = \dots = c_n \ .
\label{cdegn}
\eeq
This leaves a non-Abelian unbroken gauge group
\beq
SU(N) \longrightarrow  SU(n) \times \dots \ .
\label{nabreak}
\eeq
The first coefficient of the beta function for $SU(n)$ is
\bea
{b_0}[{SU(n)}]= (11 - 2N_{\psi}) n -6 N_{\psi} -2 N_{\chi}
 -  ( N_{\psi}+ N_{\chi}-2) (N-n)   \nn \\  =  (9-N_{\psi}+N_{\chi}) n - (6+N)N_{\psi} - (2+N) N_{\chi}+2N
\label{b0sun}
\eea
when $n=N$ this is exactly the $b_0 $ of the original $SU(N)$ theory that we choose positive (\ref{b0typeI}).   For the $\psi \chi\eta$ model where $( N_{\psi},N_{\chi}) = (1,1)$ this is always the same $\psi \chi \eta$ model reduced to $SU(n)$, and $b_0 >0$.  Where any of $N_{\psi}$ or $N_{\chi}$ is greater than $1$ we can have a sign flip for certain values of $n$. For a certain value $n^*$ we have a zero of (\ref{b0sun}).
The change of sign  for (\ref{b0sun})  happens at 
\beq
n^*= \frac{ (N_{\psi}+N_{\chi}-2)N+6N_{\psi}+2N_{\chi}}{9-N_{\psi}+N_{\chi}} \ .
\eeq
We take $[n^*]$ the biggest integer smaller than $n^*$. If we than $n$ in (\ref{cdegn}),(\ref{nabreak}) to be $[n^*]$ we have the biggest possible non-Abelian IR free sub-group.  We can see 
that when $N_{\psi}$ or $N_{\chi}$ is greater than $1$ we  have $[n^*] \simeq \frac{ (N_{\psi}+N_{\chi}-2)}{9-N_{\psi}+N_{\chi}} N $  which is greater than one and a fraction of  $N$.

\begin{figure}[h!]
\begin{center}
\includegraphics[width=6in]{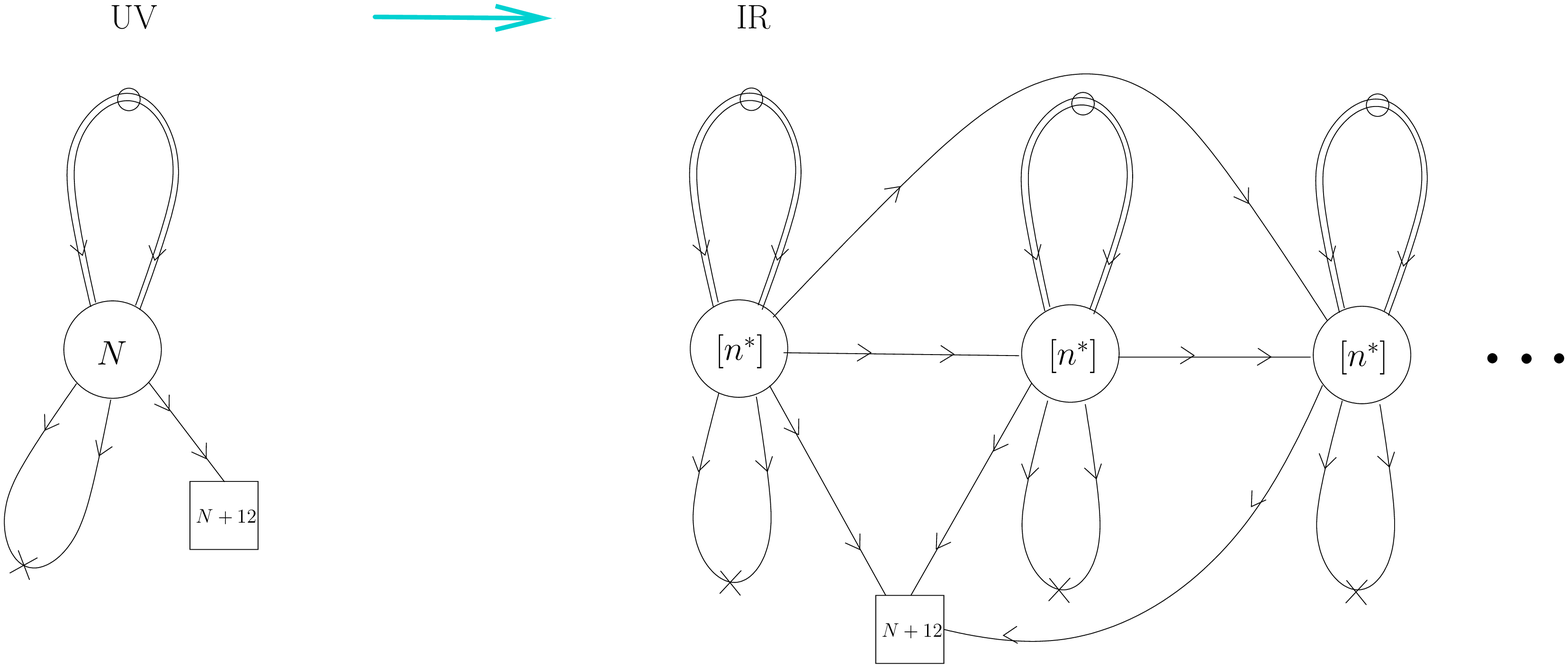}
\caption{\small  Example  of type (I) is $ (N_{\psi},N_{\chi})  = (2,1)$. }
\label{21example}
\end{center}
\end{figure}

Example  of type (I) is $ (N_{\psi},N_{\chi})  = (2,1)$: 
\be  2  \,  \yng(2)   \oplus   {\bar  {\yng(1,1)} }   \oplus (N+12)  \,  {\bar  {\yng(1)} }  \ .   \ee
This is the $\psi\eta$ model combined with the $\psi\chi\eta$ model.
We have 
\bea
b_0[{SU(N)}] = 
7N -14
\eea
$N\geq 3$,
and the first coefficient of the beta function for $SU(n)$
\bea
{b_0}[{SU(n)}] =7  n - 14 -  (N-n)    =  8 n -N -14 \ .
\eea
The change of sign happens at 
\beq
n^*= \frac{ N+14}{8} \ .
\eeq
\beq
 \begin{tabular}{|c|c|c | }
\hline
$N$&$[n^*]$&$SU(N) \rightarrow \cdots$  \\
 \hline
 $3$&$2$& $SU(3) \rightarrow SU(2) \times U(1)$ \\
$4$&$2$& $SU(4) \rightarrow SU(2) \times SU(2) \times U(1)$ \\
$5$&$2$& $SU(5) \rightarrow SU(2)  \times SU(2) \times U(1)^2$ \\
$6$&$2$& $SU(6) \rightarrow SU(2) \times SU(2)  \times SU(2) \times U(1)^2 $ \\
\vdots &\vdots & \vdots  \\
$N \to \infty$&$[n^*]$&$SU(N) \rightarrow \prod_{1}^7 SU([n^*]) \times SU(N-7 [n^*])  \times U(1)^7 $  \\

\hline
\end{tabular}  
\eeq
In Figure \ref{21example} the quiver diagrams showing the RG flow from UV to IR are shown.

\begin{figure}[h!]
\begin{center}
\includegraphics[width=6in]{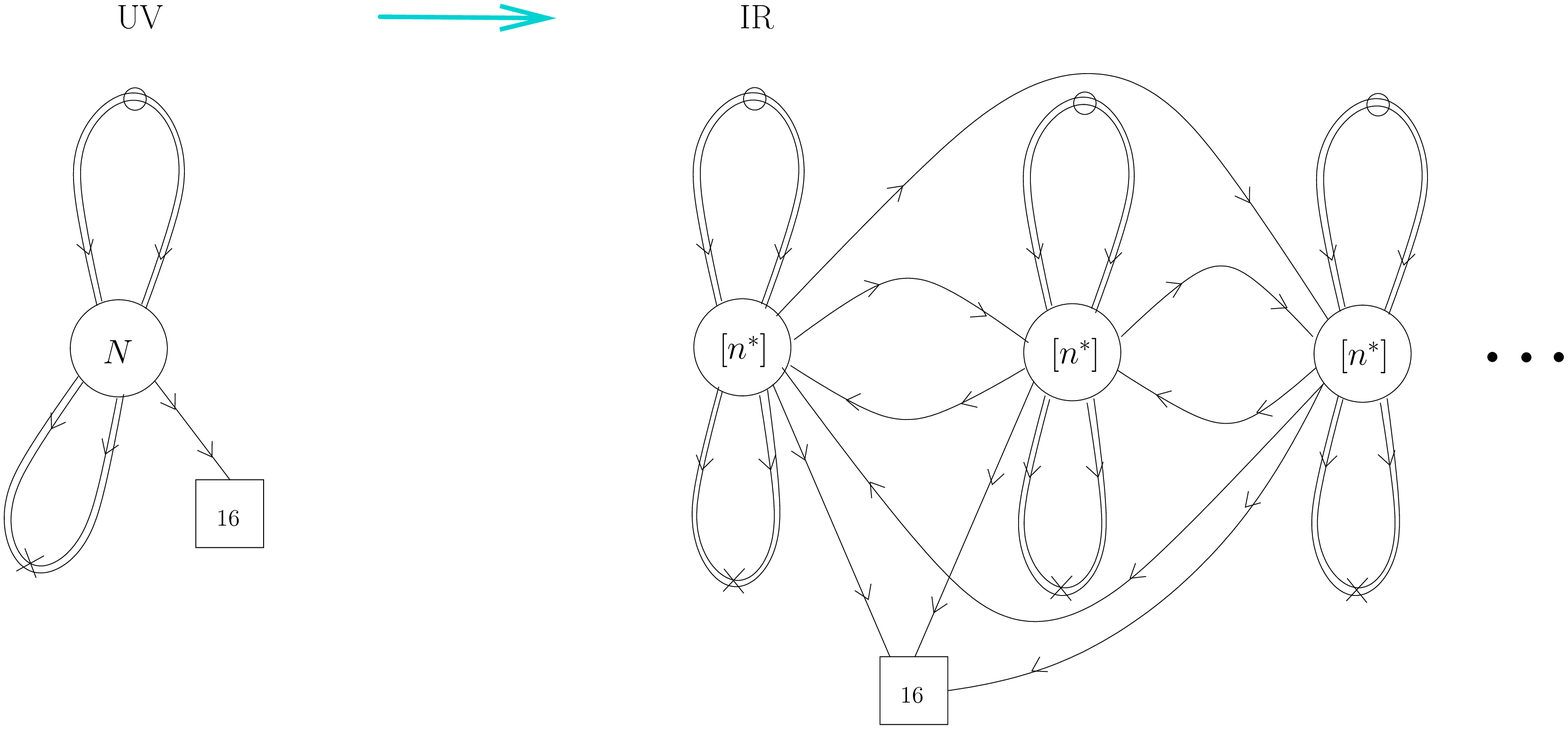}
\caption{\small  Example  of type (I) is $ (N_{\psi},N_{\chi})  = (2,2)$. }
\label{22example}
\end{center}
\end{figure}

Another example of type (I) is $ (N_{\psi},N_{\chi})  = (2,2)$: 
\be  2 \,  \yng(2)   \oplus  2  \, {\bar  {\yng(1,1)} }   \oplus  16  \,  {  \bar   {\yng(1)} }  \ ,   \ee
This is two-family version  of the $\psi\chi\eta$ model.
\bea 
b_0[{SU(N)}] 
=  7N-16
\eea
$N\geq 3$.
For $(2,2)$ 
the beta function of $SU(n)$ is
\bea
{b_0}[{SU(N)}]  =7  n - 16 - 2 (N-n)    =  9 n -2 N  -16
\eea
The change of sign happens at 
\beq
n^*= \frac{2 N+16}{9} \ .
\eeq 
\beq
 \begin{tabular}{|c|c|c | }
\hline
$N$&$[n^*]$&$SU(N) \rightarrow \cdots$  \\
 \hline
 $3$&$2$& $SU(3) \rightarrow SU(2) \times U(1)$ \\
$4$&$2$& $SU(4) \rightarrow SU(2) \times SU(2) \times U(1)$ \\
$5$&$2$& $SU(5) \rightarrow SU(2)  \times SU(2) \times U(1)^2$ \\
$6$&$3$& $SU(6) \rightarrow SU(3) \times SU(3)    \times U(1) $ \\
\vdots &\vdots & \vdots  \\
$N \to \infty$&$[n^*]$&$SU(N) \rightarrow \prod_{1}^4 SU([n^*]) \times SU(N-4 [n^*])  \times U(1)^4 $  \\

\hline
\end{tabular}  
\eeq
In Figure \ref{22example} the quiver diagrams for the RG flow  from UV to IR are illustrated.

A type (II) theory has fields
\bea   && \qquad \qquad \qquad \qquad \quad  \psi^{\{ij\},a}\;, \qquad  \chi_{[ij]}^b\;, \qquad    \tilde\eta_i^c\;,    \nn \\ &&   a =1,\ldots N_{\psi}  \;, \quad \  b=1,\ldots N_{\chi} \;, \quad  \  c =1,\ldots , N_{\chi}(N-4)  -N_{\psi}(N+4)  \;.  
\eea
We assume that only one the $\psi$'s pairs   condenses with one of the $\chi$'s. 
The $\psi\chi$ condensate breaks the global symmetry as 
\bea
&& SU(N_{\psi}) \times SU(N_{\chi}) \times SU(N_{\chi}(N-4)  -N_{\psi}(N+4)  ) \times U(1)^2  \nn  \\
&& \longrightarrow SU(N_{\psi}-1) \times SU(N_{\chi}-1) \times SU(N_{\chi}(N-4)  -N_{\psi}(N+4)  ) \times U(1)^3 \;.
\eea
We assume that the condensate can be brought to  a diagonal form like before (\ref{psichicondna}) and that there a a block of $n$ coefficients which are degenerate like (\ref{cdegn})
 This leaves a non-Abelian unbroken gauge group as (\ref{nabreak}).
The first coefficient of the beta function for $SU(n)$ is
\bea
{b_0}[{SU(n)}]=   (11 - 2N_{\chi}) n +2 N_{\psi} +6 N_{\chi}
 -  ( N_{\psi}+ N_{\chi}-2) (N-n)   \\  =  (9+N_{\psi}-N_{\chi}) n + (2-N)N_{\psi}+ (6-N) N_{\chi}+2N
\eea
when $n=N$ this is exactly the $b_0 $ of the original $SU(N)$ theory   (\ref{betafn})  that we choose positive.   For the $\psi \chi\eta$ model where $( N_{\psi},N_{\chi}) = (1,1)$ this is always the same $\psi \chi \eta$ model reduced to $SU(n)$, and $b_0 >0$.  Where any of $N_{\psi}$ or $N_{\chi}$ is greater than $1$ we can have a sign flip for certain values of $n$. For a certain value $n^*$ we have a zero of (\ref{b0sun}).
The change of sign  for (\ref{b0sun})  happens at 
\beq
n^*= \frac{ (N_{\psi}+N_{\chi}-2)N-2N_{\psi}-6N_{\chi}}{9+N_{\psi}-N_{\chi}}
\eeq

An example of type (II) is $ (N_{\psi},N_{\chi})  = (1,2)$:  
\be      \yng(2)   \oplus  2 \,  {\bar  {\yng(1,1)} }   \oplus (N-12) \,    {   {\yng(1)} }  \ ,   \ee
\bea b_0 = 11 \cdot N -   (N+2) -  2  (N-2) - (N-12)   =  7N+14
\eea
$N\geq 12$. This for $N=12$ is the $\psi\chi$ model with $k=8$. All $\psi\chi$ models are particular cases of this class.
For $12$
the beta function of $SU(n)$ is
\bea
{b_0}[{SU(n)}] =7  n + 14 -  (N-n)    =  8 n -N +14
\eea
The change of sign happens at 
\beq
n^*= \frac{ N-14}{8} \ .
\eeq

\section{Strong anomaly  \label{StrAn}}

{

Up to now we discussed the IR dynamics of the theories focusing on the realization of their global symmetry group. However these theories, at the classical level, possess another independent $U(1)$ symmetry, broken by the ABJ anomaly,  due to the nontrivial topological effects of the strong gauge interactions.    The effect is sometimes simply called as the ``strong anomaly".  
As is well-known in QCD  (the $U_A(1)$ problem and its solution)  \cite{Witten:1980sp,Witten:1979vv,DiVecchia:1980yfw,Rosenzweig:1979ay,Kawarabayashi:1980dp,Nath:1979ik} and in chiral theories \cite{Veneziano:1981yz}\cite{BKL5}, it is actually useful to keep track of  these symmetries, i.e.,  trying to reproduce their breaking in the IR theory appropriately.
Let us  briefly review how the story goes in QCD, before moving to the chiral theories of our interest. 

In QCD the axial $U(1)_A$ symmetry is broken both   by the ABJ anomaly and by the chiral condensate
\be       \brc {\bar {\psi}_L}  \psi_R \ckt    \sim -  \Lambda^3  \ne 0\;.
\ee
If one imagines the anomaly as a small explicit breaking on top of the spontaneous breaking due to the condensate, then one should have, alongside to the pions, another NGB, with a small mass due to the anomalous breaking. This scenario is quantitatively correct in the large $N$ limit.

Supposing for the moment that the anomalous breaking is small, one can describe it directly in the chiral Lagrangian,
\be  L_0=    \frac{F_{\pi}^2}{2}     \Tr   \,   \de_{\mu}  U   \de^{\mu}  U^{\dagger}     +    \Tr  M \, U  +{\rm h.c.} +   \ldots \;, \qquad  U \equiv  {\bar \psi}_R  \psi_L
\ee
by adding   a strong-anomaly effective Lagrangian   \cite{Witten:1980sp,Witten:1979vv,DiVecchia:1980yfw,Rosenzweig:1979ay,Kawarabayashi:1980dp,Nath:1979ik} 
\be       {\hat L}  =     \frac{i}{2}  q(x)  \, \log \det  U/ U^{\dagger}    +\frac{N}{a_0 F_{\pi}^2}  q^2(x)  - \theta \, q(x)\;,\label{QCDanomeff} 
\ee   
$q(x)$ is the topological density 
\be q(x)  =  \frac{g^2}{32\pi^2}  F_{\mu\nu}^a  {\tilde F}^{a, \mu\nu}\;,      \label{topodens}
\ee
$a_0$ is a constant of the order of unity,  $F_{\pi}$ the pion decay constant, and $\theta$ is the QCD vacuum parameter.  
The $U(1)_A$ anomaly under 
\be     \Delta  S  =   2 N_{\rm f}  \alpha  \int  d^4x   \frac{g^2}{32\pi^2}  F_{\mu\nu}^a  {\tilde F}^{a, \mu\nu}\;,  \qquad \psi_L \to e^{i \alpha} \psi_L\;, \quad   \psi_R \to e^{- i \alpha} \psi_R\;,
\ee
is reproduced by the  $ \log \det  U/ U^{\dagger}$ term of the effective action. 
Treating $q(x)$ as an auxiliary field, and integrating, one gets another form of the anomaly term:
\be   {\hat L}  =    -  \frac{F_{\pi}^2  \, a_0 }{4 N}   ( \theta -   \frac{i}{2}  \log \det U/U^{\dagger})^2\;. \label{analog}
\ee
 The multivalued function 
$ {\hat L}$    is actually   well defined because  
\be      \brc U \ckt  \propto {\mathbf 1}    \ne 0\;.
\ee
Moreover, as promised, this breaking is small at large $N$, so the idea of treating the anomalous breaking as a  perturbation to the chiral Lagrangian is a consistent one.

Expanding (\ref{analog}) around this VEV,
\be    U \propto  e^{  i   \tfrac{ \pi^a  t^a}{F_{\pi}}   + i    \tfrac{ \eta \, t^0}{F_{\pi}^{(0)}  }   }  =    {\mathbf 1} +    i   \frac{ \pi^a  t^a}{F_{\pi}}   + i    \frac{ \eta \,  t^0}{F_{\pi}^{(0)}  }  +\ldots \;,    \label{sigmaU}
\ee
one finds the mass term for the would-be  NG boson, $\eta$.   In real-world QCD the mass of this pseudo-NGB is large, but one can still  identify it  with the   $\eta$  meson  (in the two-flavored QCD)  or  with the $\eta'$ meson   (in the three-flavored QCD).

This discussion on the solution of the so-called $U(1)_A$  problem is well known.  An observation which might not be as familiar, is that   
the logic of the argument may be reversed: 
one can actually  argue that the presence of an effective action (\ref{analog})   needed for reproducing the strong anomaly  {\it   implies} a nonvanishing condensate,   $\brc U \ckt =   \brc \bar \psi_R \psi_L \ckt \ne 0$, and hence,  indirectly,  also  the spontaneous breaking of {\it  nonanomalous}   chiral symmetry itself,   
\be    SU(N_f)_L\times  SU(N_f)_R\times U(1)_L\times  U(1)_R      \to    U(1)_V\;,  
\ee
affecting the low-energy physics.

In the following we discuss how  the anomalous and nonanomalous  $U(1)$  (and other)  symmetries can be correctly reproduced in terms of a low-energy  effective action,    
in the $\psi\eta$,  ${\tilde \chi}\eta$,   and in  
 $\psi\chi\eta$ models. 
}

\subsection{  $\psi\eta$  model   \label{stronganomchieta}}

{
Let us apply these ideas  to the chiral gauge theories we are interested in here.
 The simplest chiral gauge theory we studied from this viewpoint  \cite{BKL5}   is the $\psi\eta$ model.

Let us review briefly  the symmetries of the model. In UV we have as global symmetry group $SU(N+4)\times U(1)_{\psi\eta}$:
any combinations of $U(1)_\psi\times U(1)_\eta$ different form $U(1)_{\psi\eta}$ is broken by a strong anomaly.  We wish to describe this anomalous breaking in the low-energy effective Lagrangian. It is particularly convenient to chose the independent anomalous   $U(1)$   symmetry as 
\be
U(1)_{an}  :\ \begin{cases}
	\psi\rightarrow e^{i\alpha}\psi\;,\\
	\eta \rightarrow e^{-i\alpha}\eta\;.\label{eq:u1a}
\end{cases}
\ee
We note that the anomalous divergence 
	\be
	\partial_\mu J^\mu_{an} = \left\{(N+2)-(N+4)\right\}\frac{g^2}{32 \pi^2} tr[F_{\mu\nu}F^{\mu\nu}]=-2\frac{g^2}{32 \pi^2} tr[F_{\mu\nu}F^{\mu\nu}]\;,
	\ee
	scales as $\mathcal{O}(1)$ for $N\to \infty$, suggesting that, for large N, a situation similar to the one in QCD (with large $N$ but fixed $N_f$): the anomalous breaking is a small effect.

Let us assume the color-flavor locked phase, discussed in Sec.~\ref{DHiggs}, characterized by the bifermion condensate $\brc \psi \eta\ckt  \sim \Lambda^3$  in (\ref{cflocking}).   
 The system breaks the gauge and part of the global symmetries dynamically,
\be
SU(N)_c\times SU(N+4)_\eta \times U(1)_{\psi\eta} \rightarrow SU(N)_{cf}\times SU(4)_f \times \tilde U(1)\;.
\ee
In order to understand the low-energy effective action, one must first identify correctly all the NG bosons. 
From the symmetry breaking one expects to find $8N$ nonAbelian NG bosons relative to   $\frac{SU(N+4)}{SU(N)\times SU(4)\times U(1)_D}$, interpolated in a gauge invariant fashion by
\be
\phi^A=(\psi^{ij}\eta_j^a)^*(T^A)^a_b (\psi^{ik}\eta^b_k)\;.
\ee
Here $T_A$ are the $8N$ broken generators that connect   the $N$ dimensional subspace (where $SU(N)$ acts)   and the $4$ dimensional one (where $SU(4)$ acts).

The Abelian symmetries in this model are slightly subtle.   The non-anomalous but spontaneously broken  
$U(1)_{\psi\eta}$ symmetry  (which implies a massless NG boson)    and anomalous (but unlike $U(1)_A$ in QCD,  not-spontaneously-broken)   $U(1)_{an}$ symmetry, 
must both correctly be reproduced in the low-energy effective action, analogous to the strong-anomaly effective action of QCD, (\ref{analog}).   As  almost any combination of $U(1)_{\psi\eta}$ and $U(1)_{an}$ is broken both by the condensate and by the anomaly one expects to have a massive (would be NG-) boson, alongside an exactly massless physical NG boson. The problem is that it is not straightforward to identify  what kind of interpolating fields, constructed from $\psi$ and $\eta$ fields describe correctly the low-energy effective action, satisfying such requirements.  Note that in the case of QCD,  the flavor-singlet combination in   $U= {\bar  \psi_R} \psi_L$ describes indeed the massive
would-be NG boson, $\eta$ (or $\eta^{\prime}$), see (\ref{analog}) and (\ref{sigmaU}).  
  
We start from the UV theory,  
\be   {\cal L}  = -  \frac{1}{4} F_{\mu\nu} F^{\mu\nu}  +      {\cal L}^{\rm fermions}   \;,
\ee
\be     {\cal L}^{\rm fermions}   
= -i \overline{\psi}{\bar {\sigma}}^{\mu}\left(\partial +\calR_{\rmS}(a)   \right)_{\mu}  \psi\;  
- i   \overline{\eta}{\bar {\sigma}}^{\mu} \left(\partial +  \calR_{\rmF^*}(a) \right))_{\mu}   \eta\;, \label{naivepsipeta}
\ee
where $a$ is the $SU(N)$ gauge field, and  the matrix representations appropriate for $\psi$ and $\eta$ fields are indicated with $\cal R_{\rmS}$ and $\cal R_{\rmF^*}$. 
To match with the IR effective Lagrangian it is useful to perform a change of variable
\be   {\cal L}  = -  \frac{1}{4} F_{\mu\nu} F^{\mu\nu}  +      {\cal L}^{\rm fermions}   +  \Tr [(\psi\eta)^* U]    +{\rm {\rm h.c.}   } 
+   { B}\,(\psi\eta\eta)^* +{\rm{\rm h.c.}}\;,    \label{above} 
\ee
where  $U$  is the composite scalars of $N\times (N+4)$ color-flavor mixed matrix form,
\be      \Tr [(\psi\eta)^* U]   \equiv    (\psi^{ij}  \eta_j^m)^*  U^{im}  \;,
\ee
and $ { B}$  are the baryons   $ B \sim \psi \eta \eta$,  
\be     B^{mn}=    \psi^{ij}  \eta_i^m  \eta_j^n \;,\qquad  m,n=1,2, \ldots, N+4\;,     \label{baryons00}
\ee
antisymmetric in  $m  \leftrightarrow n$.   In writing down the Lagrangian (\ref{above})   we have anticipated the fact that these baryon-like composite fields, present in the Higgs phase together with the composite scalars $~\psi\eta$,    are also needed to write  down the 
strong-anomaly effective action. 

Integrating $\psi$ and $\eta$ out, one gets    
\be   {\cal L}^{\rm eff}  = -  \frac{1}{4} F_{\mu\nu} F^{\mu\nu}  + \Tr  ({\cal D}  U)^{\dagger}    { \cal D} U     -i  \overline{ B}   \, { \bar {\sigma}}^{\mu}   \partial_{\mu} { B}      -  V\;.
\ee
The potential  $V$ is assumed to be such that its minimum is of the form:   
\be  \brc  U^{im}  \ckt = \, c_{\psi \eta} \,\Lambda^3  \delta^{i m }\;,   \qquad \quad \     i, m=1,2,\dots  N\;,     \label{condU}  
\ee
and contains the strong anomaly term,  
\be   V  =   V^{(0)} +    {\hat L}_{\rm an}\;.
\ee
${\hat L}_{\rm an}$ of the form,
\be       {\hat L}_{\rm an}=   { \const}  \,   \left[  \log   \left(  \epsilon \,  { B}  { B}\, { \det }  U \right)    -  \log \left(  \epsilon \, { B}  { B}  \det  U \right)^{\dagger}  \right]^2\;  \label{strong} \ee
which is analogue of (\ref{analog}) in QCD. 
The argument of the logarithm 
\be \epsilon  \, { B}  { B} \,  \det  U  \equiv    \epsilon^{m_1,m_2,  \ldots, m_{N+4}} \epsilon^{i_1, i_2, \ldots, i_N}
{ B}_{m_{N+1}, m_{N+2}}  { B}_{m_{N+3}, m_{N+4}}       U_{i_1 m_1}  U_{i_2 m_2} \ldots  U_{i_N m_N}\;, \label{insteadof}
\ee
is invariant under the full (nonanomalous) symmetry,
\be  SU(N)_{\rm c}\times SU(N+4)\times   U(1)_{\psi\eta}\;
\ee
as it should be. Moreover it contains $N+2$ $\psi$'s and $N+4$   $\eta$'s, the correct numbers of the fermion zeromodes in the instanton background: it corresponds to  a  't Hooft's instanton n-point function, e.g.,  
\be    \brc  \psi\eta\eta(x_1) \psi\eta\eta(x_2) \psi\eta(x_3) \ldots    \psi\eta(x_{N+2}) \ckt\;.
\ee

This effective Lagrangian is well defined only if the argument of the logarithm takes a VEV. In particular it is natural to assume
\be      \brc  \epsilon^{(4)}   { B}  { B}   \ckt \ne 0\;, \qquad   \brc    \det  U  \ckt  \ne 0\;,   \label{condenseBB}
\ee
where 
\be    \epsilon^{(4)}   { B}  { B}   =    \epsilon_{\ell_1  \ell_2  \ell_3  \ell_4}   { B}^{\ell_1 \ell_2}   { B}^{\ell_1 \ell_2}\;, \qquad 
\ell_i =  N+1,\ldots, N+4\;.      \label{asabove} 
\ee
As  
\be    \brc    \det  U  \ckt     \propto  {\mathbf 1}_{N\times N}\;
\ee
takes up all flavors up to $N$   (the flavor $SU(N+4)$ symmetry can be used to orient the symmetry breaking this way),  
$  { B}  { B}  $ must be made of  the four remaining flavors,  as  in (\ref{asabove}). 
These baryons were not among those considered in the earlier studies \cite{Appelquist:2000qg,BKL4}, but  assumed to be massless here,
and indicated as $B^{[A_2 B_2]}$  in Table~\ref{SimplestBis}. This is possible because these fermions do not have any perturbative anomaly with respect to the unbroken symmetry group, $SU(N)\times SU(4)\times U(1)$: 't Hooft anomaly matching considerations cannot tell if they are massive or not, either of the two options is  possible.

Now we see how the apparent difficulty about the NG bosons  hinted at  above  is solved. We can define the interpolating fields of the two NG bosons  by expanding the condensates, 
\bea    && \det  U  =  \brc \det  U\ckt +  \ldots   \propto      {\mathbf 1}  +      \frac{i}{F_{\pi}^{(0)}}  \,  \phi_0   +\ldots \;;      \nonumber \\
&& \epsilon^{(4)}   { B}  { B}   =   \brc  \epsilon^{(4)}   { B}  { B} \ckt +  \ldots  \propto
{\mathbf 1}   +     \frac{i}{F_{\pi}^{(1)}}  \,  \phi_1 +\ldots  \;, 
\eea
(here
$F_{\pi}^{(0)}$ and 
$F_{\pi}^{(1)}$  are some constants with dimension of mass). Clearly in general the physical NG boson and the anomalous would-be NG boson will be interpolated by {\it two}   linear combinations of $\phi_0$ and $\phi_1$. The effective Lagraingian determines these linear combinations.

Indeed, the effective Lagrangian  (\ref{strong})   is invariant under the nonanomalous symmetry group, and in particular $U(1)_{\psi\eta}$. 
 The boson which appears in the strong-anomaly effective action as the fluctuation of $\epsilon BB \; \det U$, 
\be      {\tilde \phi} \equiv   N_{\pi}   \left[ \frac{1}{F_{\pi}^{(0)}}   \,   \phi_0 +   \frac{1}{F_{\pi}^{(1)}}  \,  \phi_1 \right] \;, \qquad   N_{\pi} =  \frac  { F_{\pi}^{(0)} F_{\pi}^{(1)}}{
	\sqrt{\big(F_{\pi}^{(0)}\big)^2 +  \big(F_{\pi}^{(1)}\big)^2 }}  \;, 
\ee
cannot be the massless physical one: it is the would-be NG boson relative to the anomalous symmetry. Indeed the effective action provides a mass term for this would-be NG boson. 

The orthogonal combination   
\be      {\phi} \equiv   N_{\pi}   \left[ \frac{1}{F_{\pi}^{(1)}}   \phi_0   -    \frac{1}{F_{\pi}^{(0)}}  \phi_1 \right] \;,  \label{physical}  \ee 
is the interpolating field of the physical NG boson,  remaining massless.

Before we have included in the low-energy description some massless baryons which are neither required nor excluded by the 't Hooft anomaly matching. Now one can see their ultimate fate using the strong-anomaly effective Lagrangian. In particular (\ref{strong}) contains a 4-fermion coupling between these baryons, which, plugging the VEVs  (\ref{condenseBB}), provides a mass term for them. This realizes the general expectation that the system chooses the smallest amount of massless matter needed to satisfy 't Hooft anomaly matching, in line with $a$-theorem and the ACS condition.

In the above we have assumed that the system is in dynamical Higgs phase  (as discussed in Sec.~\ref{DHiggs})  and we have found out how the nonanomalous and anomalous  symmetries of the model is correctly described in the low-energy effective action.   As emphasized in \cite{BKL5},  it does not seem to be possible to write down a
 strong-anomaly effective action, if  one assumed that the system in the hypothetical, confining, symmetric vacuum, reviewed in Sec.~\ref{Hypo}.  The reason is that by using only the presumed massless degrees of freedom of the low-energy theory  (the baryons (\ref{baryons101}))  it is not possible to saturate all the fermion zeromodes needed to write down a local Lagrangian transforming appropriately under the anomalous $U(1)$ transformation.

}

\subsection{$\chi\eta$  model   \label{stronganomchieta}}

It is an interesting exercise to apply the same reasoning about the strong anomaly  to the $\chi\eta$ model. 
We will find that there are good analogies with the $\psi\eta$ case studied above, but also quite significant differences.
In this model any combination of $U(1)_\chi$ and $U(1)_\eta$, except $U(1)_{\chi\eta}$, is anomalous, therefore some term similar to (\ref{analog})   (in QCD) should appear.

In the dynamical Higgs scenario for the $\chi\eta$ model,  there are two bi-fermion condensates, 
\be  \brc  \chi^{ij} \eta_j^m  \ckt  = c_{\chi\eta}\, \delta^{im} \, \Lambda^3 \;, \qquad  i,m =  1,2,\ldots, N-4\;,   \label{chietacond}
\ee
and 
\be  \brc  \chi \chi \ckt \ne 0\;.
\ee
This implements a two-step  breaking, 
\bea        SU(N) \times    SU(N-4) \times U(1)_{\chi\eta}  &\xrightarrow{ \brc  \chi \eta \ckt}&     SU(N-4)_{\rm cf} \times  SU(4)_{\rm c} \times  U(1)^{\prime}
\nonumber \\   &\xrightarrow{ \brc  \chi \chi \ckt}& 
SU(N-4)_{\rm cf} \times  U(1)^{\prime}\;.    \label{symmetrychieta}
\eea
As before, in order to construct a fully consistent effective action, one should keep the full invariance of the original theory,  either spontaneously broken or not.
 To do so, it is convenient to re-express the condensates  (\ref{chietacond})  in a gauge invariant way.   The answer is 
\bea      U &=&    \epsilon_{i_1 i_2 \ldots i_N}    \epsilon_{m_1 m_2 \ldots m_{N-4}}    (\chi \eta)^{i_1 m_1}    (\chi \eta)^{i_2 m_2} \ldots 
(\chi \eta)^{i_{N-4} m_{N-4}}    \chi^{i_{N-3}i_{N-2} }  \chi^{i_{N-1}i_{N} } \nonumber \\
&\sim&   \epsilon \, (\chi\eta)^{N-4} (\chi\chi)\;,   \label{naturalchieta} 
\eea
which encodes both types of  the two possible composite scalar fields.

This choice suggests that   the correct  strong-anomaly  effective action  for the $\chi\eta$ model  is
\be         \frac{i}{2}    q(x)      \log    \epsilon  (\chi\eta)^{N-4} (\chi\chi)  + {\rm h.c.}\;,       \label{stronganomalychieta}    \ee    
where, again, $q(x)$ is the topological density defined in Eq.(\ref{topodens}). Clearly it  is by construction invariant under the whole (nonanomalous) symmetry group
\be 
SU(N)_{\rm c} \times  SU(N- 4) \times U(1)_{\chi\eta}\;.
\ee

This anomaly effective action  (\ref{stronganomalychieta}) agrees with  the one  proposed  by Veneziano \cite{Veneziano:1981yz} for the case of  $SU(5)$,   and 
generalizes it  to all $SU(N)$ $\chi \eta$ models. A key observation we share with\cite{Veneziano:1981yz} and generalizes to models with any $N$, is that this strong anomaly effective action,  which should be there  in the low-energy theory to reproduce correctly the (anomalous and nonanomalous) symmetries of the UV theory,  {\it  implies}   nonvanishing condensates, 
\be     \brc \chi\eta \ckt\ne 0\;, \qquad  \brc \chi\chi \ckt  \ne  0\;,     
\ee         
i.e.,  that the system is in dynamical Higgs phase.

Up to now the story has been very similar to the one about the $\psi\eta$ model. However there are some differences.
Differently from the $\psi\eta$ model, where the baryon condensate must enter in the strong-anomaly effective action, here the structure of the effective action is simpler, and no baryonlike composites are needed.
Moreover, contrary to the $\psi\eta$ model,  the $\chi\eta$  system has no physical $U(1)$ NG boson: it is eaten by a color $SU(N)$ gauge boson.    However  
the counting of the broken and unbroken $U(1)$ symmetries  is basically similar in the two models.  Of the two nonanomalous 
symmetries ($U(1)_{\rm c}$ and $U(1)_{\chi\eta}$), a combination remains a manifest physical symmetry, and the other becomes the longitudinal part of a color   gauge boson.  Still another, anomalous,  $U(1)$ symmetry exists, which is any combination of $U(1)_{\chi}$ and  $U(1)_{\eta}$ other than  
$U(1)_{\chi\eta}$.  This symmetry is also spontaneously broken,  hence it  must be associated with a NG boson, even  though it will get mass by the strong anomaly.

As in the $\psi\eta$ model, one can describe this situation  explicitly, by expanding  the composite  $ \chi\eta$ and $\chi\chi$  fields around their VEV's,   
\bea    && (\det U)^{\prime} =  \brc   (\det U)^{\prime}  \ckt +  \ldots   \propto      {\mathbf 1}  +  i   \frac{1}{ F_{\pi}^{(0)}}  \, \phi_0^{\prime}    +\ldots \;;      \nonumber \\
&& \chi \chi   =   \brc  \chi \chi  \ckt +  \ldots  \propto 
{\mathbf 1}   + i  \frac{1}{ F_{\pi}^{(1)}}   \,   \phi_1^{\prime}   +\ldots  \;, 
\eea
where $ (\det U)^{\prime} $  is defined in the $N-4$ dimensional color-flavor mixed space, and  
\be   \chi\chi  \equiv     \epsilon_{i_1,i_2,i_3,i_4} \chi^{i_1 i_2} \chi^{i_3 i_4}\;, \qquad    N-3  \le  i_j  \le  N\;. 
\ee
Now  one can see that  the strong-anomaly effective action  (\ref{stronganomalychieta})    gives mass to  
\be      {\tilde \phi}^\prime    \equiv   N_{\pi}   \left[ \frac{1}{ F_{\pi}^{(0)}}     \phi_0^{\prime}  +   \frac{1}{ F_{\pi}^{(1)}} \phi_1^{\prime}  \right] \;, \qquad   N_{\pi} =  \frac  { F_{\pi}^{(0)} F_{\pi}^{(1)}}{
	\sqrt{(F_{\pi}^{(0)})^2 +  (F_{\pi}^{(1)})^2 }}  \;, 
\ee
whereas an orthogonal combination 
\be      {\phi}^\prime    \equiv   N_{\pi}   \left[ \frac{1}{ F_{\pi}^{(1)}}     \phi_0^{\prime}  -    \frac{1}{ F_{\pi}^{(0)}} \phi_1^{\prime}  \right] \; \ee
remains massless. The latter corresponds to the  potential NG boson which is absorbed by the color  $T_{\rm c}$  gauge boson.

\subsection{   $\psi\chi\eta$  model   \label{stronganomchieta}}

  It turns out to be quite an instructive exercise to study the IR effective action of the $\psi\chi\eta$ model.    On the one hand, 
 the strong anomaly manifests differently in the dynamically Abelianized $\psi\chi\eta$ system than in the $\psi\eta$ or ${\tilde \chi}\eta$ 
  models studied above. Also, the way the nonanomalous but spontaneously broken  
 $U(1)_{\psi\chi}$  symmetry is  realized in the IR reveals some unusual, interesting features,  on the other.  
	
	As explained in Sec.~\ref{DA}, the condensate $<\psi\chi>$ breaks the global and the gauge symmetry group
	\beq     
	SU(N)_{\rm c} \times SU(8)_{\rm f} \times {\tilde U}(1)  \times  U(1)_{\psi\chi}  \longrightarrow  \prod_{\ell=1}^{N-1}  U(1)_{\ell}  \times  SU(8)_{\rm f} \times {\tilde U}(1)\;.
	\label{scenario3}
	\eeq
There are three $U(1)$ symmetries in the model \cite{BKLDA},   two non-anomalous ones  ${\tilde U}(1)$  and  $U(1)_{\psi\chi}$ 
	and  an anomalous one $U(1)_{\rm an}$. It is convenient, to decouple the different effects we want to exhibit, to take the anomalous one to be 
	\be
	U(1)_{an}: \begin{cases}
		\psi  \rightarrow  e^{i\gamma}\psi\;,\\
		\chi  \rightarrow  e^{-i\gamma}\chi\;,\\
		\eta  \rightarrow \eta\;.
	\end{cases}
	\ee  
The condensate  (\ref{psichicond})  does not break it.
	
	The associated currents are
	\bea
	&&J_{\psi\chi}^{\mu} =   i     \left\{     \tfrac{N-2 }{N^* }   \,   {\bar \psi}  {\bar \sigma}^{\mu}   \psi -        \tfrac{N+2 }{N^* } \, {\bar \chi}  {\bar \sigma}^{\mu}   \chi  \right\} \;,
	\qquad  \de_{\mu}   J_{\psi\chi}^{\mu}  = 0\;,    \label{current1}   \\
	&&  \ {\tilde J}^{\mu}   =     i  \,   \Big\{    2  \, {\bar \psi}  {\bar \sigma}^{\mu}   \psi -     2  \, {\bar \chi}  {\bar \sigma}^{\mu}   \chi   -      {\bar \eta^a}  {\bar \sigma}^{\mu}   \eta^a  \Big\}  \;,
	\qquad \  \   \   \de_{\mu}    {\tilde J}^{\mu}   = 0\;,  \phantom{\frac{1}{2}}   \label{current2}\\
	&& \, J_{\rm an}^{\mu}  =      i  {\bar \psi}  {\bar \sigma}^{\mu}   \psi -    i  {\bar \chi}  {\bar \sigma}^{\mu}   \chi   \;,   \qquad  \qquad  \qquad  \qquad  \   \ \  \,  \de_{\mu}   \,  J_{\rm an}^{\mu}  =     \frac{ 2 g^2}{32 \pi^2}   G_{\mu \nu}  {\tilde G}^{\mu \nu}  \;. 
	\label{current3}
	\eea 
	Among $U(1)_{an},\;\tilde U(1)\,U(1)_{\psi\chi}$, only $U(1)_{\psi\chi}$ is broken by the condensate, one expects a single NG oson, $\pi$. At this point one can start to write down an effective Lagrangian
	\be    {\cal L}^{(eff)} =   {\cal L}(\psi, \eta,  A_{\mu}^{(i)} )  +   {\cal L}(\pi)    -    {\cal V}(\pi,  \psi, \eta)    + \ldots  \;,   \label{effective}   \ee
	with $\pi$, the photons, $A^k$, $k=1, ..., N-1$, and the massless fermions, $\psi^{ii}, \eta^A_i$. Now we can use $U(1)_{an}$ and $U(1)_{\psi\chi}$ to learn as much as possible about ${\cal L}^{(eff)}$.
	
	$U(1)_{an}$ is an (anomalously) broken symmetry, therefore  the IR lagrangian  must  break it. In terms of the UV fields, the anomalous conservation equation leads to the  't Hooft vertices, 
	\be    \epsilon_{a_1 a_2 \ldots  a_8}   \underbrace{   (\psi\chi)^i_j    (\psi\chi)^j_k \ldots     (\psi\chi)^p_i }_{N-2}   \underbrace{  (\psi \eta^{a_1}  \eta^{a_2})  \ldots  ( \psi  \eta^{a_7}   \eta^{a_8} )}_{4}\;,    \label{thooft}  
	\ee
	where a possible (certainly not unique)  way to contract the color $SU(N)$ and the flavor $SU(8)$  indices in an invariant way is shown. 
	By defining $U(x)$
	\be    U(x) =    (\psi\chi)^1_1(x)  =    \const. \,\Lambda^3   \, e^{i \pi(x)  / F}       \label{Ufield} 
	\ee 
	one can capture this contribution in the IR effective theory by including 
	\be   {\cal V}(\pi,  \psi, \eta)(x)     \sim  U(x)^{N-2}  \, \underbrace {\psi \cdots \psi}_{4}  \, \underbrace {\eta \, \eta \cdots \eta}_{8}   + h.c.   =   \const. \,    \pi \pi \ldots \pi \; \, \underbrace {\psi \cdots \psi}_{4}\,  \underbrace {\eta\, \eta \cdots \eta}_{8}  + h.c.   \;,       \label{infra}  
	\ee
	with  any number of   pions,  four $\psi$'s and eight $\eta$'s.   This term is invariant under the full UV symmetries     $SU(N)_{\rm c} \times SU(8)_{\rm f} \times {\tilde U}(1)  \times  U(1)_{\psi\chi} $, 
	and {\it a fortiori}     with the unbroken symmetries $ \prod_{\ell=1}^{N-1}  U(1)_{\ell}  \times  SU(8)_{\rm f} \times {\tilde U}(1)$, but clearly not under the anomalous  $U(1)_{an}$.  
	
	The effect of the $U(1)_{an}$   anomaly, however,   is not exhausted in the explicit breaking of  $U(1)_{an}$  symmetry in  $ {\cal V}(x)$. 
	As   the  $U(1)_{an}$  charge of the low-energy, massless fermions is well defined,  
	\be
	\psi^{ii} \rightarrow e^{i\gamma}\psi^{ii}\;, \quad \eta^A_i \rightarrow \eta^A_i\;,
	\ee  
	it  manifests itself   also  through  the massless  $\psi$  fermion loops,  
	\be        J_{\rm an}^{\mu}  =      i \sum_{i=1}^N   {\bar \psi}_{ii}   {\bar \sigma}^{\mu}   \psi^{ii}  \;, \qquad    
	\de_{\mu}   J_{\rm an}^{\mu}  =     \frac{1}{16 \pi^2}  \sum_{j=1}^{N-1}     e_{j}^2  \, F_{\mu \nu}^{(j)}    {\tilde F}^{(j)\,   \mu \nu}  \;.\label{U1anomalies}
	\ee
	Such an anomaly has a   natural interpretation as a remnant of the original strong anomaly  (\ref{current3})   in the UV theory.     The original strong anomaly divergence equation has turned into the anomalous divergences due to the weak  $U(1)^{N-1}$ gauge interactions of the low-energy theory. 
	
	Let us  now  turn our attention to $U(1)_{\psi\eta}$. This is an exact symmetry of the UV theory, so it must be (although non-linearly realized) symmetry in the IR theory. In particular, in IR, this becomes the shift symmetry for the NGB, 
	\be        \pi(x)  \to    \pi(x)  -  \frac{4 F}{N^*}   \beta \;,\qquad U(x) \to  e^{-  \tfrac{4 i \beta}{N^*} }   \, U(x)\;,   \label{piontr}  
	\ee
	but still act non-trivially on    $\psi^{ii}$,  $i=1,2,\ldots, N$:
	\be          \psi^{ii}      \to  e^{i  \tfrac{N-2 }{N^* }    {\beta}} \psi^{ii}\;.   \label{psitr} 
	\ee
	But now  the  anomaly due to the   $\psi^{ii}$ loops,   
	\be       \Delta   {\cal L}^{eff}  =     \frac{N-2}{N^*} \beta     \frac{1}{16 \pi^2}  \sum_{j=1}^{N-1}     e_{j}^2  \, F_{\mu \nu}^{(j)}    {\tilde F}^{(j)\,   \mu \nu}    \label{psianomaly}
	\ee
	is not cancelled, leading to the apparent paradox:  $U(1)_{\psi\chi}$   is a nonanomalous (exact) symmetry of the system, but in the low-energy effective theory 
	it seems to be broken by anomaly!	
The answer to this puzzle is that   the low-energy  effective Lagrangian (\ref{effective}) contains an axion-like term 
	\be           {\cal L}(\pi,  A_{\mu}^{(i)} )    =      \pi(x) \,       \frac{N-2}{4 F}  \frac{1}{16 \pi^2}  \sum_{j=1}^{N-1}     e_{j}^2  \, F_{\mu \nu}^{(j)}    {\tilde F}^{(j)\,   \mu \nu} 
	\label{axionlike} 
	\ee
	which transforms under   (\ref{piontr})   as 
	\be       \Delta   {\cal L}(\pi,  A_{\mu}^{(i)} )     =    -     \frac{N-2}{N^*} \beta     \frac{1}{16 \pi^2}  \sum_{j=1}^{N-1}     e_{j}^2  \, F_{\mu \nu}^{(j)}    {\tilde F}^{(j)\,   \mu \nu}  
	\ee
	cancelling exactly the anomaly due to the   $\psi^{ii}$ loops,  (\ref{psianomaly}), ensuring the    $U_{\psi\chi}(1)$ invariance of the system.

\section{Conclusion \label{Concl}}

We considered in this work  a large class of simple SU(N) chiral gauge theories which admit large N limit, some of which are relatively unexplored.   
	We discussed possible IR effective theories, realization of symmetries and matching of anomalies.
	In some limited class of models,  conventional 't Hooft anomaly matching constraints alone,  appears to suggest that it is possible that 
	the system flows into a confinement phase, without global symmetry breaking,  i.e., with no condensates forming. However, the considerations
	of tighter constraints following from the generalized symmetries and mixed anomalies \cite{BKL2,BKL4}, as well as those based on the strong anomaly \cite{BKL5}, 
	clearly disfavor such a dynamical possibility. 	
	Dynamical Higgs mechanism, instead,  is a very natural solution, and appears to be consistent.   It is also consistent with the mixed-anomaly 
	matching as well as with the strong anomaly constraints.  A particular variation of dynamical Higgs phase, which might occur in some chiral gauge theories
	with composite bifermion scalar field in the adjoint representation of $SU(N)$,  is dynamical Abelianizatio, discussed in a separate section  (Sec.~\ref{DA})  above.  
	
Which is the “real” channel of condensation is a hard dynamical question. 
A maximally attractive channel (MAC) idea  suggests that  the condensate form in the most strongly attractive bi-fermion composite state. But such a
na\"ive consideration based on the one-gluon exchange picture cannot tell which is the right answer, when  
there are several possible competing condensates having similar attractive strength.  Dynamical arguments would be required to go further.\footnote{
 Some attempts have been made recently, by starting from a supersymmetric  version of the  models, and by adding a small 
 susy-breaking term as perturbation.  In some cases, they show that a sort of dynamical Higgs  phase is realized \cite{Csaki:2021aqv,Csaki:2021xhi,Bai:2021tgl}. It is however not easy to extend these results  when supersymmetry breaking becomes large. In particular,  the most bifermion condensates appearing in the dynamical Higgs
 phase, discussed in Sec.~\ref{DHiggs}  here, are forbidden by supersymmetry,   meaning that the dynamical Higgs phases in these chiral theories cannot be reached from the supersymmetric version of the models, by susy-breaking perturbation. 
  }
Another line of development is the  adiabatic continuity to a semiclassical regime using compactification. This was started for chiral theories in \cite{Shifman:2008cx} and very recently applied to the $\psi\chi\eta$ model \cite{Sheu:2022odl}. The  result  \cite{Sheu:2022odl} is consistent with dynamical Abelianization  \cite{BKLDA}.

 We considered effective actions of massless fermions and NB bosons.
 We studied the different types of realization of strong anomaly in the IR. 
	There are peculiar features of chiral theories regarding discrete symmetries and generalized symmetries we have not discussed here (see \cite{BKLReview} for a review).

An interesting open problem for the future is to write the effective action for all cases, and especially for the ones with non-Abelian symmetry breaking patterns in  IR.

An interesting line for future developments is to study the soliton in the effective Lagrangian.  In large $N$ QCD,  the Skyrmions of the chiral Lagrangian are identified with the baryon.  Their mass scales like $N$ and the WZW provides the baryon charge to the soliton. The same happens in (S)QCD and (A)QCD with heavy baryons made of $\frac{N(N\pm1)}{2}$ quarks \cite{Bolognesi:2006ws}. 
Large $N$ solitons  and heavy baryons should appear also in these chiral theories, and hopefully they would be related in some way to the effective actions we constructed.

The hope is that some of these theories will turn out to be useful in the context of realistic model building.  
 The Glashow-Weinberg-Salam (GWS) $SU(2)_L\times U(1)_Y$  theory (as well as its GUT generalizations) is a weakly coupled theory and,  as such, is well defined and well understood within the  perturbation theory  framework. 
But this also means that the 
theory should be regarded, at best, as a very good low-energy effective theory.  In particular, 
it is unlikely that  the gauge symmetry breaking sector 
described by a potential term for the Higgs scalar, though phenomenologically quite successful, is a self-consistent, fundamental description. 
Nevertheless,  attempts to replace it by new, QCD-like strongly-coupled gauge theories 
(technicolor, extended technicolor,  walking technicolor, etc.)  
have not been fully successful so far.  At the same time, our rather limited understanding of {\it strongly-coupled chiral gauge theories} has been hindering us from
making a concrete progress by using them so far.   We hope that our new proposals involving the low-energy effective 
non-Abelian gauge symmetries could find a phenomenological application in near future.

\section*{Acknowledgments} 

  The work is supported by the INFN special 
research initiative grant,  “GAST" (Gauge and String Theories).

\section*{References}

\bibliography{ProceedingFinalFinal} 
\bibliographystyle{ieeetr}

\end{document}